\documentclass[a4paper,11pt]{article}
\usepackage{amsmath,amssymb,amsthm,amsxtra,overpic,bbm,bbold,bm,epsfig,multirow}
\usepackage{color,ulem,subfigure,hyperref,cite,float}

\allowdisplaybreaks

\begin{document}
\begin{titlepage}
\begin{center}
{\bf\Large
\boldmath{
Exploring flavour space of an economical SU(5) GUT in future proton decay measurements}
}
\\[12mm]
Gao-Xiang~Fang$^{1,2,3}$\footnote{E-mail: \texttt{fanggaoxiang21@mails.ucas.ac.cn}}
and
Ye-Ling~Zhou$^{1,}$\footnote{E-mail: \texttt{zhouyeling@ucas.ac.cn}}
\\[-2mm]
\end{center}
\vspace*{0.50cm}
\centerline{$^{1}$\it School of Fundamental Physics and Mathematical Sciences,}
\centerline{\it Hangzhou Institute for Advanced
Study, UCAS, Hangzhou 310024, China}
\centerline{$^{2}$\it Institute of Theoretical Physics, Chinese Academy of Sciences, Beijing 100190, China}
\centerline{$^{3}$\it University of Chinese Academy of Sciences, Beijing 100049, China}

\vspace*{1.20cm}

\begin{abstract}
{\noindent
We discuss the potential of future proton decay experiments on the exploration of the flavour space of grand unification. We focus on an economical $SU(5)$ grand unified model (GUT) with the fermion sector extended by including only one copy of 24-plet. Neutrino masses are generated via type-(I+III) seesaw mechanism with the lightest neutrino massless. Gauge unification requires masses of fermions in the 24-plet to be hierarchical, in particular, the electroweak singlet and triplet heavy leptons to be around the canonical seesaw scale and  TeV scale, respectively. We address how extra parameters in the flavour space which cannot be touched in flavour measurements can be tested by a multi-channel analysis in future proton decay measurements. 
}
\end{abstract}
\end{titlepage}

\section{Introduction}

Grand unification has always been the pursuit of physicists, as it can naturally unify three fundamental forces into a single force and unveil new physics beyond the Standard Model (SM). Grand unified theory (GUT) utilizes the power of symmetries to embed the SM gauge group $SU(3)_C\times SU(2)_L\times U(1)_Y$ in a simple gauge group. The simplest one is proposed by H. Georgi and S. Glashow in 1974, based on SU(5) gauge group \cite{Georgi:1974sy}. Since quarks and leptons are embeded in common irreducible representations of the GUT group, it predicts the decay of proton, which can be mediated by GUT gauge bosons \cite{Pati:1973rp}. This impressive prediction makes GUT models in principle testable by current and future experiments.
	
It is well-known that the Georgi-Glashow model has been ruled out due to the prediction of massless neutrinos, wrong mass relations between down-type quarks and charged leptons, and the failure of gauge couplings \cite{Georgi:1974yf}. Two distinctive ways have been proposed to save this theory. One effective way is to utilize higher-dimensional  operators to correct above problems, especially dimension 5 operators, seeing \cite{Ellis:1979fg,Weinberg:1979sa,Shafi:1983gz} and very recently \cite{Senjanovic:2024uzn}. Another is to introduce more fields in the approach of model building, causing many minimal extended $SU(5)$ theories on the market. One can generate neutrino masses through different seesaw mechanisms. By introducing a new symmetric  Higgs representation ${\bf 15}_H$ \cite{Glashow:1979nm,Senjanovic:1981ff}, the scalar triplet $({\bf 1},{\bf 3},1)_\Delta$ is responsible for generating neutrino masses through type-II seesaw mechanism \cite{Magg:1980ut,Mohapatra:1980yp,Lazarides:1980nt}. Alternative approach is to add an additional fermionic representation ${\bf 24}_F$ \cite{Bajc:2006ia,Bajc:2007zf} instead of ${\bf 15}_H$. The fields responsible for generating neutrino masses through type-I and -III seesaw mechanisms are $({\bf 1}, {\bf 1}, 0)_N$ and $({\bf 1}, {\bf 3}, 0)_\Sigma$. An interesting  prediction is the weak triplet fermions should be light enough to increase unification scale, maybe within the reach of LHC \cite{Bajc:2006ia}.
Further developments of these extensions are found in \cite{Dorsner:2005fq,Dorsner:2006hw,Dorsner:2007fy,DiLuzio:2013dda,Dorsner:2014wva,FileviezPerez:2016sal,Hagedorn:2016dze,Dorsner:2019vgf,Antusch:2021yqe,Antusch:2022afk,Antusch:2023kli,Antusch:2023mqe}.
	
As the most dramatic prediction of grand unified theories, proton decay is a powerful window to test if a GUT model is allowed by current experimental limits and within the ability of future experiments. Typical channels $p \to \pi^{0}e^{+}$ and $p \to K^{+}\bar{\nu}$ have been constrained in the measurement in Super-Kamiokande (SK) experiments with the best lower bound of proton partial lifetime $\tau_{\pi^{0}e^{+}}>2.4\times 10^{34}$ years \cite{Super-Kamiokande:2020wjk}, $\tau_{K^{+}\bar{\nu}}>6.6\times 10^{33}$ years \cite{Takhistov:2016eqm} at 90\% confidence level (CL). Furthermore, future experiments will push the sensitivity of proton lifetime measurement to a higher precision level, e.g., $\tau_{K^{+}\bar{\nu}}>9.6\times 10^{33}$ years in JUNO \cite{JUNO:2022qgr}, $\tau_{K^{+}\bar{\nu}}>1.3\times 10^{34}$ years in DUNE \cite{DUNE:2020ypp}, and $\tau_{\pi^{0}e^{+}}>7.8\times 10^{34}$ years and $\tau_{K^{+}\bar{\nu}}>3.2\times 10^{34} $ years in Hyper-Kamiokande (HK) \cite{Hyper-Kamiokande:2018ofw}. Another channels will  also be measured up to a higher sensitivities \cite{Dev:2022jbf}.
	
Following the above discussion, one can think about different realistic extensions of the Georgi-Glashow model. In this article, we construct a simple renormalizable extension of the Georgi-Glashow model that corrects the three major problems by including extra fermionic  ${\bf 24}_F$ multiplets. 	
This article is organised as follows. In Section~\ref{sec:2}, we construct the framework of this model and derive mass matrices of particles. In Section~\ref{sec:3}, we realise gauge coupling unification and scan the parameter space of new particles, obtain the range of unification scale. In the rest of this paper, we will limit our discussion in the minimal extension, i.e., only one copy of ${\bf 24}_F$ introduced.  Section~\ref{sec:4} discusses how the light neutrino masses are generated via type-(I+III) seesaw. In Section~\ref{sec:5}, we focus on the correlation between the flavour space and partial lifetime of proton in multi decaying channels.  Comparing with current and future experimental limits, we point out targeted parameter space for future experiments to explore.

\section{The framework}\label{sec:2}

We give a brief description of the particle contents and their interactions in this section.

The SM matter fields in the $SU(5)$ gauge space are arranged in a $\bar{\mathbf{5}}$-plet and a $\mathbf{10}$-plet. They are respectively represented as rank-$(0,1)$ and rank-$(2,0)$ tensors in the space of fundamental representations, which are, in detail, written to be
\begin{eqnarray}
	\overline{\bf 5}_F =
	\left(
	\begin{array}{c}
		d_{rR}^c \\
		d_{gR}^c \\
		d_{bR}^c \\
		e_L \\
		-\nu_L
	\end{array}
	\right) \,,\quad
	%%%%%%%%%%
	{\bf 10}_F = \frac{1}{\sqrt{2}}
	\left(
	\begin{array}{ccccc}
		0 & u_{bR}^c & -u_{gR}^c & -u^r_L & -d^r_L \\
		-u_{bR}^c & 0 & u_{rR}^c & -u^g_L & -d^g_L \\
		u_{gR}^c & -u_{rR}^c & 0 & -u^b_L & -d^b_L \\
		u^r_L & u^g_L & u^b_L & 0 & -e_R^c \\
		d^r_L & d^g_L & d^b_L & e_R^c & 0
	\end{array}
	\right) \,.
\end{eqnarray}
We include one more matter field which transforms as a 24-dimensional adjoint representation of $SU(5)$,
\begin{eqnarray}
	{\bf 24}_F = \begin{pmatrix}
		& & & U_{rR}^c & D_{rR}^c \\
		& \{\lambda_a Q_8^a - \frac{2}{\sqrt{30}}N\}^i_j & & U_{gR}^c & D_{gR}^c \\
		& & & U_{bR}^c & D_{bR}^c \\
		U_{rL} & U_{gL} & U_{bL} & \frac{1}{\sqrt{2}} \Sigma^0 + \frac{3}{\sqrt{30}} N & \Sigma^+ \\
		D_{rL} & D_{gL} & D_{bL} & \Sigma^- & \frac{-1}{\sqrt{2}} \Sigma^0 + \frac{3}{\sqrt{30}} N \\
	\end{pmatrix} \,,
\end{eqnarray}
where $i,j=1,2,3$, and $\lambda_a$ for $a=1,2,...,8$ are Gell-Mann matrices. This multiplet includes a SM gauge singlet fermion $N$ and an electroweak triplet $\Sigma$. They help to generate tiny neutrino masses via seesaw mechanisms. It further includes a colour-octant fermion $Q_8$ and a pair of colour-triplet vector-like fermions $Q = (U_L, D_L)^T,  (U_R, D_R)^T$. These particles are all assumed to be heavier than the electroweak scale and have important contributions to cure the gauge unification as will be discussed in the next section. 

{In the Higgs sector, we include three Higgses, ${\bf 5}_H$, ${\bf 45}_H$ and  ${\bf 24}_H$. For former two are necessary for the decomposition to the standard model Higgs and to give different masses for the SM quarks and leptons. The last one, ${\bf 24}_H$, on one hand, is introduced to break $SU(5)$ to the SM gauge symmetries as originally suggested \cite{Georgi:1974sy}. On the other hand, it will be used to split masses for fermions in the ${\bf 24}_F$ representation, additional Higgs ${\bf 24}_\Phi$ are required. Note that introducing ${\bf 24}_{\Phi}$ is enough to split masses of fermions in ${\bf 24}_F$ \cite{FileviezPerez:2007bcw,FileviezPerez:2008afb}. This representation include singlet of the SM gauge symmetry. In this work, from the economical point of view, we will assume they are heavy enough thus only their vacuum expectation values (VEVs) contributes to the low-energy phenomenology.} 

\begin{table}[h]
	\begin{center}
		\begin{tabular}{ l c l}
			\hline \hline
			Fields & $SU(5)$ & $\supset SU(3)_c \times SU(2)_L \times U(1)_Y$  \\
			\hline
			Fermion & $\mathbf{10}_F$ & $\to ({\bf 3}, {\bf 2}, +\frac{1}{6})_{q_L} + (\overline{\bf 3}, {\bf 1}, -\frac{2}{3})_{u_R^c} + ({\bf 1}, {\bf 1}, +1)_{e_R^c}$  \\
			& $\overline{\mathbf{5}}_F$ & $\to (\overline{\bf 3}, {\bf 1}, +\frac{1}{3})_{d_R^c} + ({\bf 1}, {\bf 2}, -\frac{1}{2})_{\ell_L}$  \\
			& $\mathbf{24}_F$ & $\to ({\bf 1}, {\bf 1}, 0)_N + ({\bf 1}, {\bf 3}, 0)_\Sigma + ({\bf 3}, {\bf 2}, -\frac{5}{6})_{Q_L} + (\overline{\bf 3}, {\bf 2}, \frac{5}{6})_{Q_R^c}+ ({\bf 8}, {\bf 1}, 0)_{Q_8}$  \\
			\hline
			Higgs & $\mathbf{5}_H$ & $\to ({\bf 1}, {\bf 2}, \frac{1}{2})_{h_1}$  \\
			& ${\bf 45}_H$ & $\to ({\bf 1}, {\bf 2}, \frac{1}{2})_{h_2}$ \\
			& ${\bf 24}_\Phi$ & $\to  ({\bf 1}, {\bf 1}, 0)_{\varphi_1}$ \\
			\hline \hline
		\end{tabular}
	\end{center}
	\caption{Particle contents of $SU(5)$ and their decomposition to the SM gauge group $SU(3)_c \times SU(2)_L \times U(1)_Y$.} The subscript for each representation indicates the label of the corresponding field. In the fermion sector, both ${\bf 10}_F$ and $\overline{\bf 5}_F$ have 3 copies, which are decomposed to all fermion fields in the SM. Extra $n_f$ copies of fermion presentation ${\bf 24}_F$ are introduced to generated light neutrino mass via type-(I+III) seesaw mechanism. 
	\label{tab:particle_contents}
\end{table}

Lagrangian terms to generate down quark and charged lepton masses are
\begin{eqnarray} \label{eq:superpotential_up}
	- {\cal L}& \supset&
	\bar{\bf 5}_F (Y_1  {\bf 5}_H^\dag + Y_2 {\bf 45}_H^\dag) {\bf 10}_F + {\bf 10}_F (Y_3  {\bf 5}_H + Y_4 {\bf 45}_H) {\bf 10}_F  \nonumber\\
	&+& \bar{\bf 5}_F  (Y_5  {\bf 5}_H + Y_6 {\bf 45}_H)  {\bf 24}_F + {\bf 24}_F (M_1 + \kappa_1 {\bf 24}_\Phi) {\bf 24}_F + {\rm h.c.} \,,
\end{eqnarray}
where ${\cal C} = i \gamma^2 \gamma^0$ is the charge-conjugate matrix. The first two terms lead to Yukawa coupling matrices for down-type quarks and charged leptons, the third and fourth terms generate up-type quark masses.
In the left-right convention (i.e., $\overline{\psi_L} M_\psi \psi_R$), these mass matrices are written as
\begin{eqnarray}
	M_e &=& a Y_1^* - 3b Y_2^* \,, \nonumber\\
	M_d &=& a Y_1^\dag + b Y_2^\dag \,, \nonumber\\
	M_u &=& c Y_3^* + d Y_4^*  \,.
\end{eqnarray}
The fifth and sixth terms generate  Dirac neutrino mass between $\nu_L$ and $N$, as well as that between $\nu_L$ and $\Sigma$,
\begin{eqnarray}
	M_{\rm I} &=& \sqrt{3} f Y_5^* + \sqrt{5} g Y_6^* \,, \nonumber\\
	M_{\rm III} &=& \sqrt{5} f Y_5^* - \sqrt{3} g Y_6^* \,.
\end{eqnarray}
The last three terms give the Majorana mass for $\Sigma$ and $N$ and split their masses from $Q_8$ and $Q$.
In order to split the mass of $\Sigma$ from those of other fermions in the ${\bf 24}_F$ representation and improve the GUT scale, {we include one  representation of scalars, ${\bf 24}_{\Phi}$. It includes trivial singlet representation of $SU(3)_c \times SU(2)_L \times U(1)_Y$, which are indicated as $\varphi_1$ as shown in Table~\ref{tab:particle_contents}. The VEV of this singlet leads not just to the break of $SU(5)$ to $SU(3)_c\times SU(2)_L \times U(1)_L$, but also helps to split masses of $N$, $\Sigma$, $Q$ and $Q_8$. After the this singlet gains the VEV, all entries of ${\bf 24}_\Psi$ representation gain masses as
\begin{eqnarray}
	&&M_N=|m- e^{i\alpha} \Lambda| \,, \nonumber\\
	&&M_\Sigma= |m - 3 e^{i\alpha} \Lambda| \,, \nonumber\\
	&&M_{Q}= |m -\frac{1}{2} e^{i\alpha} \Lambda| \,, \nonumber\\
	&&M_{Q_8}= |m + 2 e^{i\alpha} \Lambda| \,,
	\label{eq:24Fmass}
\end{eqnarray}
where {$m=|M_1|$, $\Lambda = |\kappa_1 \langle \varphi_1 \rangle|$ and $\alpha$ is the relative phase between $M_1$ and $\kappa_1 \langle \varphi_1 \rangle$.} Different situation where $\alpha$ takes different specific values and corresponding results have been discussed in \cite{FileviezPerez:2008afb}.}
The light neutrino mass matrix is obtained via type-(I+III) seesaw formula
\begin{eqnarray}
	M_\nu = M_{\rm I} M_N^{-1} M_{\rm I}^T + M_{\rm III} M_\Sigma^{-1} M_{\rm III}^T \,.
\end{eqnarray}

\section{Unification of gauge couplings}

\label{sec:3}

\subsection{RG running of gauge couplings} \label{sec:gauge_unifcation}

Given the Standard Model (SM) gauge group $G_{\rm SM} = SU(3)_C \times SU(2)_L \times U(1)_Y$ with respect to gauge couplings
\begin{eqnarray} \label{eq:gauge_couplings}
	\{g_i\} = \begin{pmatrix} g_3 \\ g_2 \\ g_1\end{pmatrix} \,,
\end{eqnarray}
the renormalisation group (RG) running equation for each coupling is given by
\begin{eqnarray}
	\frac{d\alpha_i(t)}{dt} = \beta_i (\alpha_j)\,,
\end{eqnarray}
where $t = \log(\mu/\mu_0)$ and $\alpha_i =g_i^2/(4\pi)$. Here we fix $\mu_0$ at $m_Z$. The $\beta$ function at two-loop level is expressed in the form \cite{King:2021gmj}
\begin{eqnarray} \label{eq:RGE}
	\beta_i = \frac{1}{2\pi} \alpha_i^2 ( a_i + \frac{1}{4\pi} \sum_{j} b_{ij} \alpha_j ) \,.
\end{eqnarray}
Here, $a_i$ and $b_{ij}$ are the coefficients obtained from one- and two-loop corrections, respectively.
Given gauge symmetries $H_i \times H_j$, they are generically given by
\begin{eqnarray}
	a_i &=& - \frac{11}{3} C_2(H_i) + \frac{2}{3} \sum_F T(\psi_i) + \frac{1}{3} \sum_S T(\varphi_i) \,, \nonumber\\
	b_{ij} &=&
	- \frac{34}{3} [C_2(H_i)]^2 \delta_{ij} + \sum_F T(\psi_i) [2 C_2(\psi_j) + \frac{10}{3} C_2(H_i) \delta_{ij}] \nonumber\\
	&&+ \sum_S T(\varphi_i) [4 C_2(\varphi_j) + \frac{2}{3} C_2(H_i) \delta_{ij}]\,,
\end{eqnarray}
where the $\psi$ and $\varphi$ indices sum over the fermions and complex scalar multiplets, respectively, and $\psi_i$ and $\varphi_i$ are their representations in the group $H_i$, respectively.
$C_2(H_i)$ is the quadratic Casimir of the adjoint presentation of the group $H_i$, and in particular, $C_2(SU(N))=N$ and $C_2(U(1)) = 0$.
$C_2(R_i)$ (for $R_i=\psi_i, \varphi_i$) denotes the quadratic Casimir of the representation $R_i$ in group $H_i$, and
for the fundamental irrep ${\bf N}$ of $SU(N)$, $C_2({\bf N}) = (N^2-1)/2N$.
$T(R_i)$ is the Dynkin index of representation $R_i$ of group $H_i$. For $SU(N)$, the Dynkin index is expressed as $T(R_i) = C_2(R_i)d(R_i)/(N^2-1)$ where $d(R_i)$ is the dimension of $R_i$.
If $H_j = U(1)$, $b_{ij}$ should be obtained by replacing $C_2(R_j)$ and $T(R_j)$ with the charge square $[Q_j(R)]^2$ of the field multiplet $R$ in $U(1)$.

The coefficients $b_i$ and $b_{ij}$ account for degrees of freedom of fields running in loops.
Their explicit values depend on the degrees of freedom of fermions and scalars introduced in the theory.
Above the electroweak scale, all SM particles contribute as
\begin{eqnarray}
	\{a_i^{\rm SM}\} =  \left(
	\begin{array}{c}-7 \\ -\frac{19}{6} \\ \frac{41}{10}\\
	\end{array}
	\right) \,, \quad
	\{b_{ij}^{\rm SM}\} = \left(
	\begin{array}{ccc}
		-26 & \frac{9}{2} & \frac{11}{10} \\
		12 & \frac{35}{6} & \frac{9}{10} \\
		\frac{44}{5} & \frac{27}{10} & \frac{199}{50} \\
	\end{array}
	\right) \,.
\end{eqnarray}
Here, $ i=1,2,3 $ labels the three gauge interactions in $SU(3)_C$, $SU(2)_L$, and $U(1)_Y$, respectively,  following the ordering in Eq.~\eqref{eq:gauge_couplings}.
We further include contributions of extra fermions introduced from the ${\bf 24}$-plet in Table~\ref{tab:particle_contents}. Note that these fermions may have hierarchical masses thus their contributions to the RG running are scale-dependent. For each copy of fermion, the corresponding $\beta$ coefficients are given by
\begin{eqnarray} \label{beta coefficients}
	\{a_i^{\Sigma}\} =\; \begin{pmatrix}0 \\ \frac43 \\ 0 \end{pmatrix}\; \,, &\quad&
	\;\;\{b_{ij}^{\Sigma}\} = \; \left(
	\begin{array}{ccc}
		0 & 0 & 0 \\
		0 & \frac{64}{3} & 0 \\
		0 & 0 & 0 \\
	\end{array}
	\right) \,, \nonumber\\
	\{a_i^{Q}\} = \begin{pmatrix} \frac43\\ 2\\ \frac{10}{3} \end{pmatrix} \,, &\quad&
	\;\; \{b_{ij}^{Q}\} = \left(
	\begin{array}{ccc}
		\frac{76}{3} & 3 & \frac53 \\
		8 & \frac{49}{2} & \frac52 \\
		\frac{40}{3} & \frac{15}{2} & \frac{25}{6} \\
	\end{array}
	\right)\,, \nonumber\\
	\{a_i^{Q_8}\} = \; \begin{pmatrix} 2\\ 0\\ 0 \end{pmatrix}\; \,, &\quad&
	\;\{b_{ij}^{Q_8}\} = \; \left(
	\begin{array}{ccc}
		48 & 0 & 0 \\
		0 & 0 & 0 \\
		0 & 0 & 0 \\
	\end{array}
	\right) \,.
\end{eqnarray}
We emphasise that the above coefficients account for only one copy's contribution. If there are $n_f$ copies of ${\bf 24}_F$ involved above a certain scale, a factor $n_f$ should be included for all these coefficients. 
It is notable that these particles contribute only if the threshold effect is open, i.e., the scale $\mu$ larger than the mass of relevant particles.

The RG equation, Eq.~\eqref{eq:RGE}, can be analytically solved.
We first discuss the solution at the one loop level by ignoring $b_{ij}$ contributions. Gauge couplings are then solved to be
\begin{equation} \label{eq:alpha_1_loop}
	\alpha_{i}^{-1}(t)=\alpha_{i}^{-1}(0)-\frac{a_{i}^{\rm SM}}{2\pi}t-\sum_{I}\frac{a_{i}^{I}}{2\pi}(t-t_{I})\theta(t-t_{I}) \,,
\end{equation}
Here, the last term on the right hand side refers to the one-loop threshold correction induced by the new particle $I$ with its mass $M_I$ deviating from the symmetry breaking scale, where $\theta(t-t_{I})$ is the Heaviside step function and $t_I = \log(M_I/m_Z)$. In this paper, we consider only the fermion sector contributes to the threshold effect, i.e., $I = \Sigma, Q, Q_8$, and $m_Z<M_I<M_{\rm GUT}$. We assume all particles have masses between the electroweak scale and the GUT scale.
By taking $ t=t_{\rm GUT} $,  gauge couplings at the GUT scale are given by
\begin{equation}
	\alpha_{i}^{-1}(t_{\rm GUT})=\alpha_{i}^{-1}(0)-\frac{a_{i}^{\rm SM}}{2\pi}t_{\rm GUT}-\sum_{I}\frac{a_{i}^{I}}{2\pi}(t_{\rm GUT}-t_{I})  \,.
\end{equation}
As gauge couplings are unified at the GUT scale, we have tree-level matching conditions,
\begin{eqnarray}
	\alpha_{1}^{-1}(t_{\rm GUT})=\alpha_{2}^{-1}(t_{\rm GUT})=\alpha_{3}^{-1}(t_{\rm GUT})\,.
\end{eqnarray}
Once the matching condition is satisfied, we derive the GUT scale $ t_{\rm GUT} $ with respect to $ t_{I} $ as
\begin{equation}
	t_{\rm GUT}=\frac{2\pi\delta\alpha_{32}^{-1}+\sum_{I}\delta a_{32}^{I}t_{I}}{\delta a_{32}^{\rm SM}+\sum_{I}\delta a_{32}^{I}} \,.
\end{equation}

While the one-loop results help to demonstrate the analytical behaviour between different mass scales required by the gauge unification, a qualitative restriction on these scales from proton decay measurements requires the RG equation up to the two-loop level.
We take the two-loop RG running effect into account, and the solution of RG equation is then modified to
\begin{eqnarray} \label{eq:alpha_2_loop}
	\alpha_{i}^{-1}(t_{\rm GUT})&=&\alpha_{i}^{-1}(0)- \frac{\tilde{a}_{i}^{\rm SM}}{2\pi}t_{\rm GUT}
	-\sum_{I} \frac{\tilde{a}_{i}^{I}}{2\pi}(t_{\rm GUT}-t_{I}) \,,
\end{eqnarray}
where
\begin{eqnarray} \label{eq:alpha_2_loop_2}
	\tilde{a}_{i}^{\rm SM} &=& a_{i}^{\rm SM} - \sum_{j}\frac{b_{ij}^{\rm SM}\log(1-w_j(0) t_{\rm GUT})}{2 a_{j}^{\rm SM}t_{\rm GUT}} \,, \nonumber\\
	\tilde{a}_{i}^I &=& a_{i}^I - \sum_{j}\frac{b_{ij}^I\log[1-w_j(t_I) (t_{\rm GUT}-t_I)]}{2 a_{j}^{\rm SM}(t_{\rm GUT}-t_I)}  \,,
\end{eqnarray}
$w_j(t) = \alpha_{j}(t) a_{j}^{\rm SM} /(2\pi)$. As seen from the above formula, we also take two-loop threshold correction, represented by $b_{ij}^I$, into account. This effect, as later shown in Fig.~\ref{fig:RGE_running}, can be quantitatively important if some mass $M_I$ has a large deviation from the symmetry breaking scale. To a good approximation, we replace $\alpha_j(t)$ in the expression of $w_j(t)$ by the one-loop result in Eq.~\eqref{eq:alpha_1_loop}.
Furthermore, two-loop RG running equations require one-loop matching conditions for self consistency, and the latter are given by
\begin{eqnarray} \label{eq:1_loop_matching}
	\alpha_3^{-1}(t_{\rm GUT})-\frac{3}{12\pi} = \alpha_2^{-1}(t_{\rm GUT})-\frac{2}{12\pi} = \alpha_1^{-1}(t_{\rm GUT}) \,.
\end{eqnarray}

\begin{figure}[t!] 
	\begin{center} 
		\includegraphics[width=0.47\textwidth]{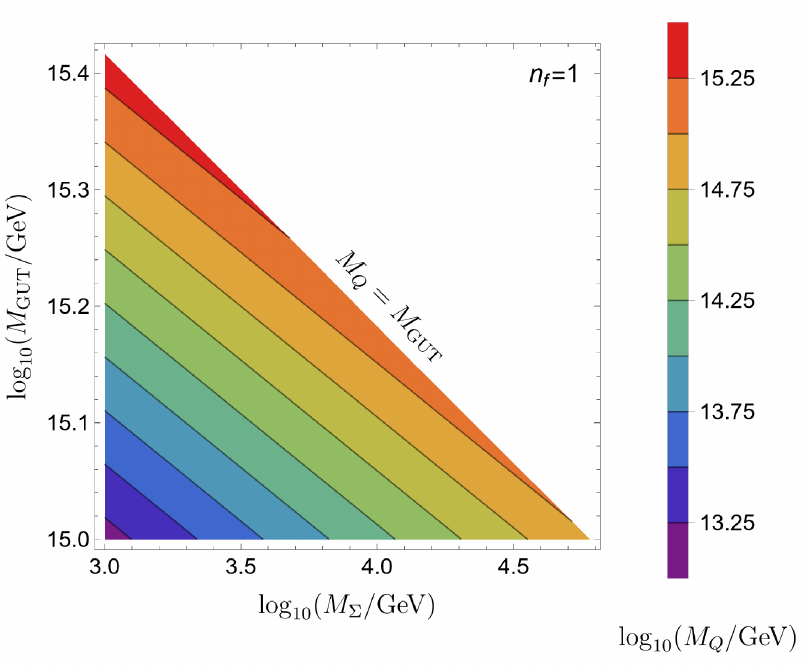}
		\includegraphics[width=0.47\textwidth]{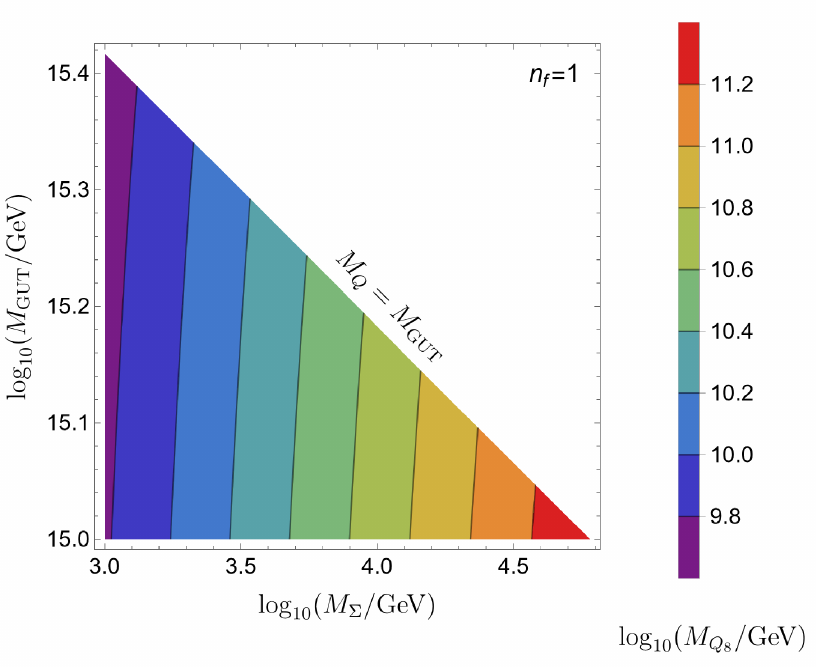}
		\includegraphics[width=0.47\textwidth]{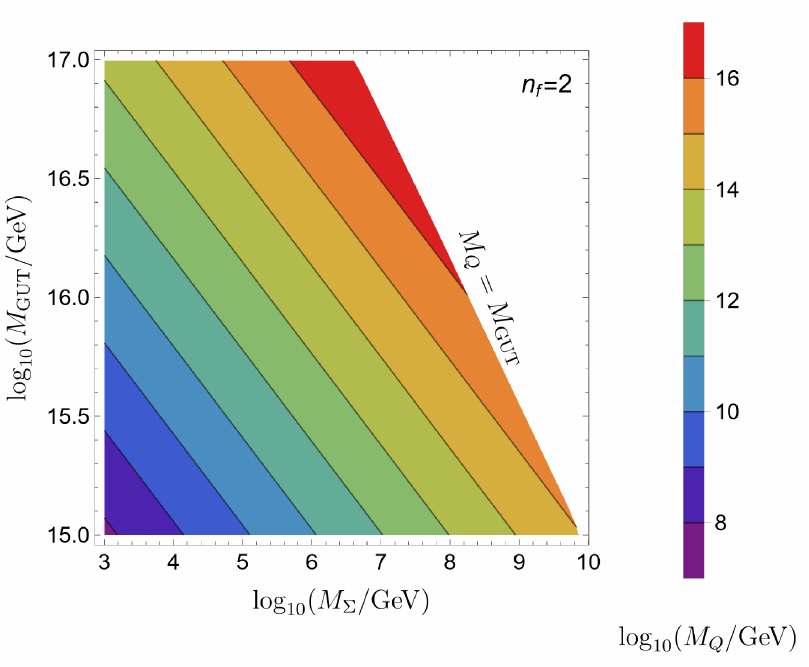}
		\includegraphics[width=0.47\textwidth]{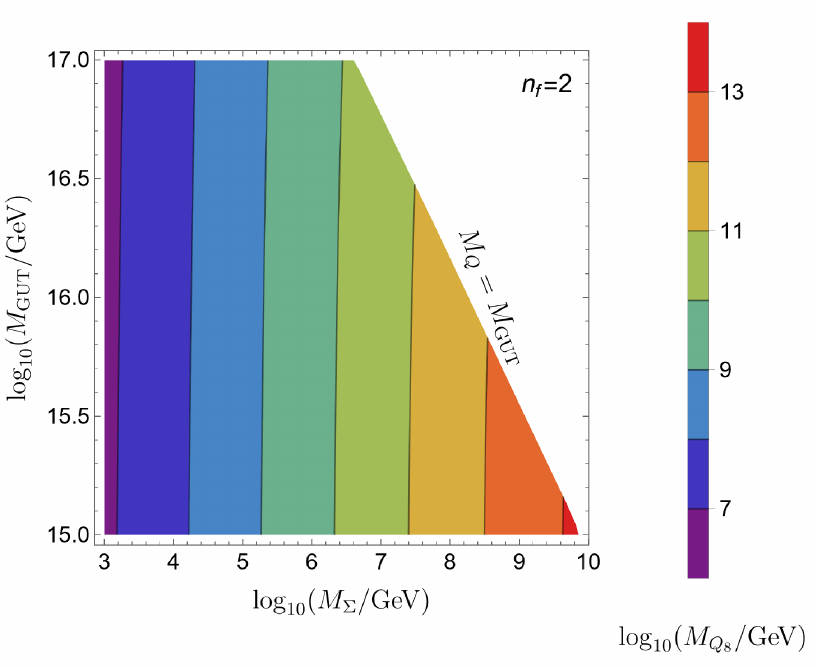}
		\includegraphics[width=0.47\textwidth]{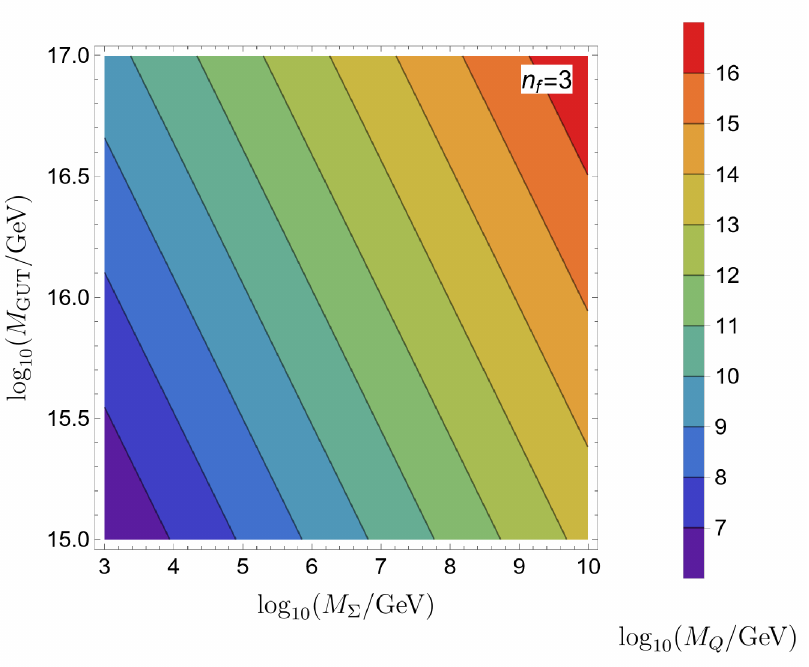}
		\includegraphics[width=0.47\textwidth]{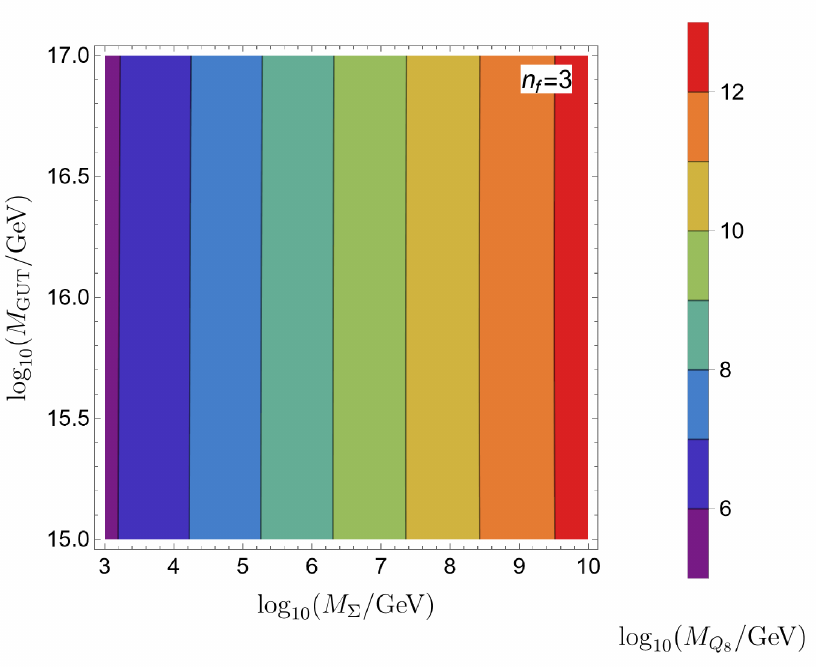}
		\caption{Correlations between heavy lepton masses and the GUT scale required by the Gauge Unification. Contours for $ M_{Q}$ (left panel) and $M_{Q_{8}}$ (right panel) are given with $ M_{\Sigma} $ and $ M_{\rm GUT} $ treated as variables. Mass ranges $ M_{\Sigma} $ in $ 10^{3} \sim 10^{10} {\rm GeV} $ and $ M_{\rm GUT} $ in $ 10^{15} \sim 10^{17} {\rm GeV}$ are considered and the mass scale hierarchy $M_{\Sigma}, M_Q, M_{Q_8} \leqslant M_{\rm GUT}$ is required. $n_f = 1, 2, 3$ denotes the number of $\mathbf{24}_{F}$'s copy. }
		\label{fig:scales}
	\end{center}
\end{figure}

\subsection{Correlation between heavy lepton masses and $M_{\rm GUT}$}
\label{sec:3.2}

In the gauge space, the SM matter fields belong to $\bar{\mathbf{5}}$- and $\mathbf{10}$-plets. In the $ SU(5) $ model, the new matter fields are included in the $\mathbf{24}$-plets, the field  decomposition from $ SU(5) $ to the SM gauge group $SU(3)_c \times SU(2)_L \times U(1)_Y$ and the particles contained in each representation are listed in Table~\ref{tab:particle_contents}. 
New particles which have contribution to RG equations are $ \Sigma, Q, Q_{8}$. We calculate the $ \beta $ coefficients of these new particles in Eq.~\eqref{beta coefficients}. Using Eq.~\eqref{eq:1_loop_matching}, we obtain the parameter space of $M_{\Sigma} $ and $ M_{\rm GUT} $ and show the dependence of $ M_{Q}$ and $M_{Q_8} $ upon them. %%
Furthermore, we will set $M_\Sigma \ge 10^3$~GeV for simplicity while performing scan. This is consistent with the lower bound of heavy leptons set by the collider searches at ATLAS ($M_\Sigma > 910$~GeV at 95\% confidence level) \cite{ATLAS:2022yhd}.
By fixing the copy of ${\bf 24}_F$ at $n_f = 1,2,3$, correlations of these scales are presented in the upper, middle and lower panels of Fig.~\ref{fig:scales}. 
Here we have assumed masses of all particles in ${\bf 24}_F$ not heavier than $M_{\rm GUT}$ such that the theory is always in the perturbative regime. 
For $n_f=2,3$, degenerate masses between different copies are assumed for illustration. 
As seen in the plots, the allowed parameter space gradually increases as the copy of ${\bf 24}_F$ increases. When ${\bf 24}_F$ has only one copy ($n_f=1$), the maximum value of $M_{\rm GUT}$ is about $2.63 \times 10^{15}$ GeV ($M_{\Sigma}=10^3$ GeV),  $M_{\Sigma}$ ranges from 1 TeV to 63 TeV under the condition where $M_{\rm GUT} \ge 10^{15}$ GeV. 
{This means a strongly fine tuning between $m$ and $\Lambda$ is required to generate a much smaller value of $M_\Sigma$.}
In the rest of this work, we only focus on the situation where ${\bf 24}_F$ has only one copy.

\begin{figure}[t!] 
	\begin{center} 
		\includegraphics[width=0.425\textwidth]{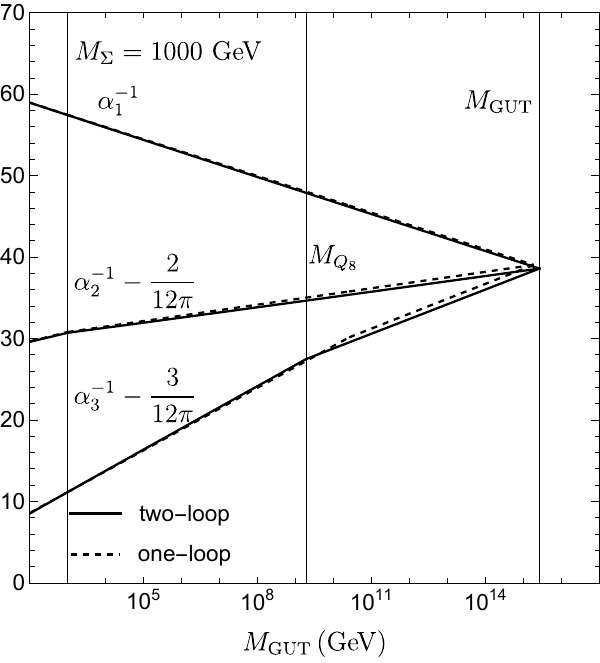}
		\includegraphics[width=0.44\textwidth]{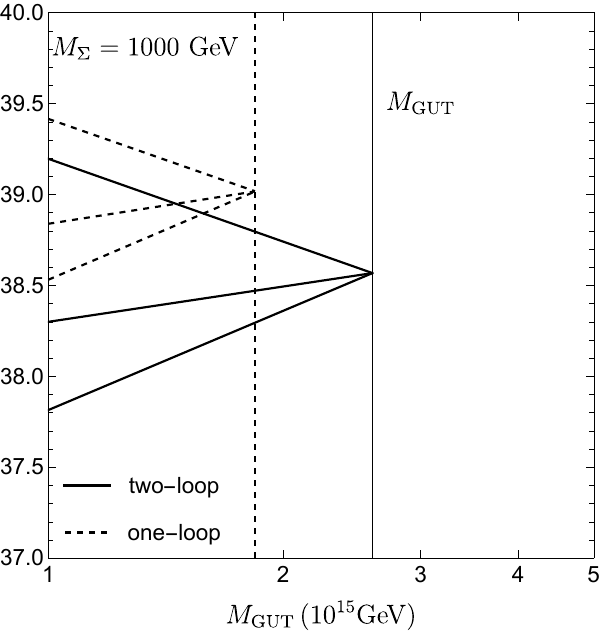}
		\caption{Running of gauge couplings at one-loop (dashed) and two-loop (solid) levels. With $M_\Sigma=10^3$ GeV and $M_Q =M_{\rm GUT}$ fixed, the GUT scale is calculated to be $M_{\rm GUT} = 1.84 \times 10^{15}$ at one-loop and $2.63 \times 10^{15}$ GeV at two-loop, respectively. }
	\label{fig:RGE_running}
	\end{center}
\end{figure}

By setting $n_f = 1$, we further show an example of gauge couplings running respectively at one- and two-loop levels in Fig.~\ref{fig:RGE_running} with $M_\Sigma = 10^3$~GeV and $M_Q = M_{\rm GUT}$ fixed. 
As seen in the figure, the two-loop RG running helps to enhance the GUT scale $M_{\rm GUT}$ by roughly 40\% compared with the one-loop running, namely, $M_{\rm GUT} = 2.63\times 10^{15}$~GeV at two-loop compared with $M_{\rm GUT} = 1.84\times 10^{15}$~GeV at one-loop. This enhancement is quantitatively important to obtain the proton lifetime because the latter is proportional to $M_{\rm GUT}^4$.

\begin{figure}[t]
	\begin{center}
		\includegraphics[width=0.6\textwidth]{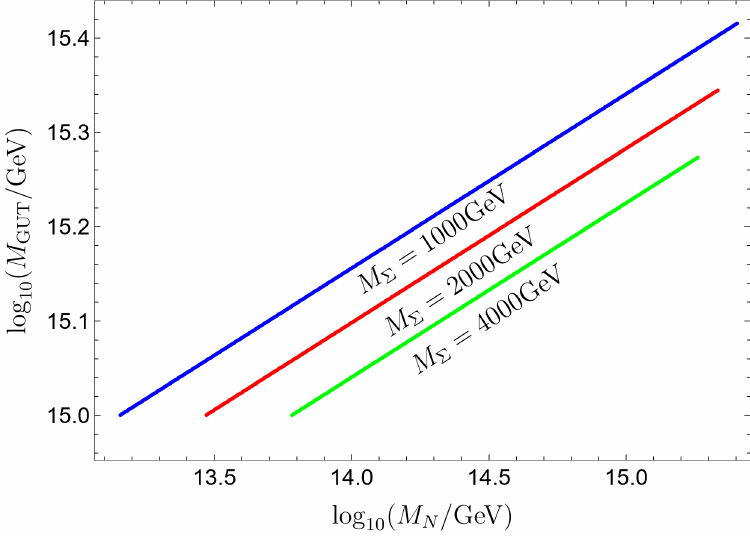}
		\caption{The relation between $M_{\rm GUT}$ and $M_{N}$ by setting $M_{\Sigma}=1, 2, 4 \times 10^3~{\rm GeV}$, respectively.}
		\label{fig:scales_2}
	\end{center}
\end{figure}

{Based on the masses of $N$, $\Sigma$, $Q$ and $Q_8$ calculated in Eq.~\ref{eq:24Fmass}, we can derive the correlation among $\Sigma$, $Q$, $Q_8$ and $N$ masses
\begin{eqnarray}
	M_{N}=\frac{1}{5}(3M_{\Sigma}^2-2M_{Q_8}^2+24M_{Q}^2)^{\frac{1}{2}} \,.
\end{eqnarray}
}Combining with correlations between heavy lepton masses and the GUT scale required by the gauge unification in Fig.~\ref{fig:scales}, $M_{N}$ is only relevant to $M_{\Sigma}$ and $M_{\rm GUT}$. Setting $M_{\Sigma}=1,2, 4\times 10^3~{\rm GeV}$, we get the relation between $M_{\rm GUT}$ and $M_{N}$, illustrated in Fig.~\ref{fig:scales_2}. 
As we can see, under the condition $M_{\rm GUT}>M_{N}$, the maximal value of $M_{\rm GUT}$ is about $2.63 \times 10^{15}$ GeV with $M_{\Sigma}=10^3$ GeV. When $M_{\rm GUT}$ ranges in $(1,2.63) \times 10^{15}$ GeV, $M_{N}$ is constrained in $(3.20\times10^{13},2.63 \times 10^{15})$ GeV. The maximal value of $M_{\rm GUT}$ and the range of $M_{N}$ gets small while $M_{\Sigma}$ gets larger. When $M_\Sigma=10^3$ GeV, the maximal value of $M_{\rm GUT}$ is $2.63 \times 10^{15}$ GeV, which is allowed by gauge unification.

\begin{figure}[t!] 
	\begin{center} 
		\includegraphics[width=0.425\textwidth]{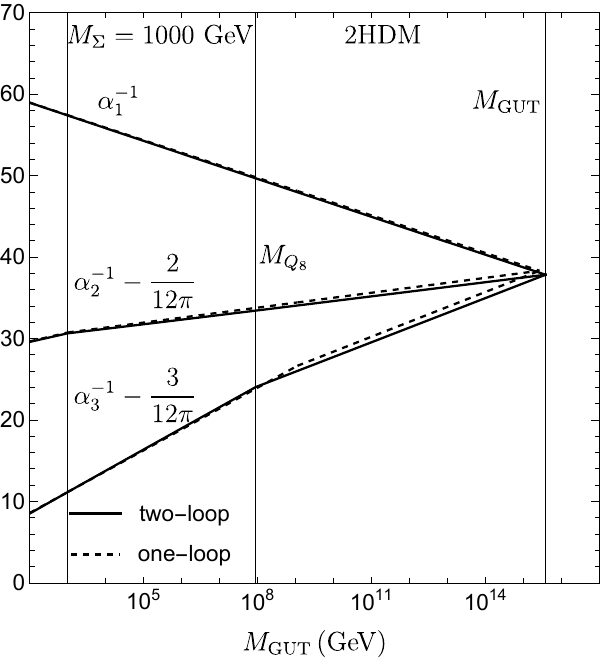}
		\includegraphics[width=0.44\textwidth]{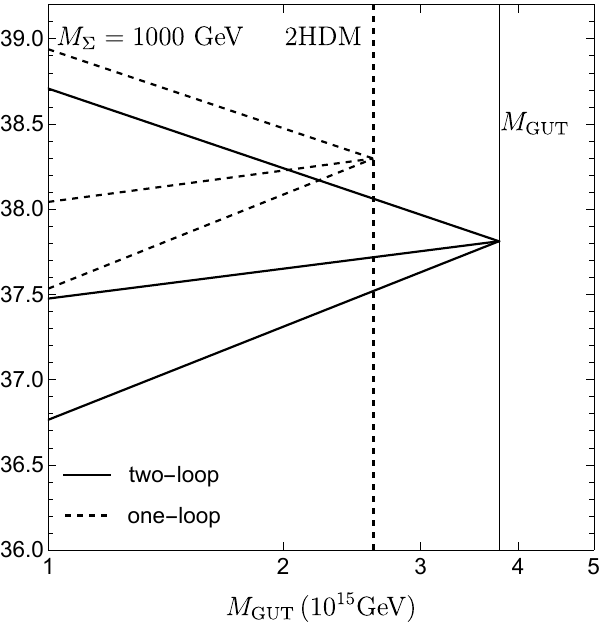}
		\caption{Running of the couplings at one-loop level and two-loop level with two-Higgs-Doublets Model (2HDM) assumed at the electroweak scale. Same inputs and setups are included as the last figure. $M_{\rm GUT}$ is calculated to be $2.61 \times 10^{15}$ and $3.78 \times 10^{15}$ GeV. }
		\label{fig:RGE_running_2HDM}
	\end{center}
\end{figure}

We check the influence of extra higgs doublet on the gauge unification. 
In our model, it it natural to have both Higgs EW-doublets decomposed from ${\bf 5}_H$, ${\bf 45}_H$ exist at scales much lower than the intermediate scale $M_\Sigma$. We consider a simple case with two-Higgs-Doublet Model (2HDM) running from the EW scale. Then, the $\beta$ coefficient including the extra doublet Higgs contribution is given by 
\begin{eqnarray}
	\{a_i^{\rm 2HDM}\} =  \left(
	\begin{array}{c}-7 \\ -3 \\ \frac{21}{5}\\
	\end{array}
	\right) \,, \quad
	\{b_{ij}^{\rm 2HDM}\} = \left(
	\begin{array}{ccc}
		-26 & \frac{9}{2} & \frac{11}{10} \\
		12 & 8 & \frac{6}{5} \\
		\frac{44}{5} & \frac{18}{5} & \frac{104}{25} \\
	\end{array}
	\right) \,,
\end{eqnarray}
which is slightly different from those in the SM case. As a comparison to Fig.~\ref{fig:RGE_running}, we show an example of RG running of gauge couplings at two-loop level in Fig.~\ref{fig:RGE_running_2HDM} in 2HDM situation, where all inputs keep the same. The maximal value of $M_{\rm GUT}$ is increased by approximately a half compared to that in Fig.~\ref{fig:RGE_running}.

\section{Fermion masses and mixing}\label{sec:4}

All charged fermion Yukawa coupling matrices can be bilinearly diagonalised by two unitary matrices, 
\begin{eqnarray}
	U_{u}\hat{Y_{u}}U_{u}^{\prime\dagger}=Y_{u},\quad U_{d}\hat{Y_{d}}U_{d}^{\prime\dagger}=Y_{d},\quad	U_{e}\hat{Y_{e}}U_{e}^{\prime\dagger}=Y_{e},
\end{eqnarray}
where a hatted matrix represents a diagonal matrix with all non-vanishing entries positive, and 
and $U_{u},\,U_{d},\,U_{e},\,U_{u}^{\prime},\,U_{d}^{\prime},\,U_{e}^{\prime}$ are unitary matrices. 
Without loss of generality, we will work in the diagonal charged-lepton flavour basis where $U_{e} = U_{e}'={\bf 1}$.

The CKM matrix is defined as $V_{\rm CKM}
\equiv U_u^\dag U_d$.
Dismissing the five unphysical phases, the CKM matrix is parametrised as
\begin{eqnarray}
	V_{\rm CKM} =
	\left(\begin{matrix}
		c^{q}_{12} c^{q}_{13} & s^{q}_{12} c^{q}_{13} &
		s^{q}_{13} e^{-{ i} \delta^{}_q} \cr \vspace{-0.4cm} \cr
		-s^{q}_{12} c^{q}_{23} - c^{q}_{12}
		s^{q}_{13} s^{q}_{23} e^{{ i} \delta^{}_q} & c^{}_{12} c^{q}_{23} -
		s^{q}_{12} s^{}_{13} s^{}_{23} e^{{ i} \delta^{}_q} & c^{q}_{13}
		s^{q}_{23} \cr \vspace{-0.4cm} \cr
		s^{q}_{12} s^{q}_{23} - c^{q}_{12} s^{q}_{13} c^{q}_{23}
		e^{{ i} \delta^{}_q} &- c^{q}_{12} s^{q}_{23} - s^{q}_{12} s^{q}_{13}
		c^{q}_{23} e^{{ i} \delta^{}_q} &  c^{q}_{13} c^{q}_{23} \cr
	\end{matrix} \right),
\end{eqnarray}
where $s_{ij}^q=\sin\theta_{ij}^{q},\,c_{ij}^q=\cos\theta_{ij}^{q}$. In the numerical calculation below, we will fix values of CKM mixing angles $(\theta)$ and CP violating phase at
\begin{eqnarray}
	\theta_{12}^{q}=0.227,\quad \theta_{23}^{q}=4.858 \times 10^{-2},\quad \theta_{13}^{q}=4.202\times 10^{-3},\quad \delta^{q}=1.207 \,,
\end{eqnarray}
where are obtained by running their experimental best-fit values to the GUT scale $M_{\rm GUT}$ \cite{Antusch:2013jca,Babu:2016bmy}.
Note that there is a large redundancy of free parameters in the flavour sector.

In the neutrino sector, the Majorana mass matrix for light neutrino is in general a symmetric and complex matrix, which can be diagonalised by a unitary matrix $U_\nu$ via,
\begin{eqnarray} \label{eq:diagonalise_M_nu}
U_\nu \hat{M}_\nu U_\nu^T = M_\nu \,.
\end{eqnarray}
The product $U_e^\dag U_\nu$ gives the PMNS matrix. The latter is parametrised as
\begin{eqnarray} \label{eq:PMNS}
	U_{\rm PMNS} =
	\left(\begin{matrix}
		c^{}_{12} c^{}_{13} & s^{}_{12} c^{}_{13} &
		s^{}_{13} e^{-{ i} \delta} \cr \vspace{-0.4cm} \cr
		-s^{}_{12} c^{}_{23} - c^{}_{12}
		s^{}_{13} s^{}_{23} e^{{ i} \delta} & c^{}_{12} c^{}_{23} -
		s^{}_{12} s^{}_{13} s^{}_{23} e^{{ i} \delta} & c^{}_{13}
		s^{}_{23} \cr \vspace{-0.4cm} \cr
		s^{}_{12} s^{}_{23} - c^{}_{12} s^{}_{13} c^{}_{23}
		e^{{ i} \delta} &- c^{}_{12} s^{}_{23} - s^{}_{12} s^{}_{13}
		c^{}_{23} e^{{ i} \delta} &  c^{}_{13} c^{}_{23} \cr
	\end{matrix} \right)\times \text{diag}\{1,e^{i\alpha_{21}/2},e^{i \alpha_{31}/2}\},
\end{eqnarray}
up to three unphysical phases.
We apply the best-fit values from NuFIT 5.3\cite{Esteban:2020cvm, nufit5.3} in the latter numerical calculation,
\begin{eqnarray}
	\theta_{12}=33.66^{\circ},\quad \theta_{23}=49.1^{\circ},\quad \theta_{13}=8.54^{\circ},\quad \delta=197^{\circ},\nonumber\\ \Delta m_{21}^{2}=7.41\times10^{-5}\, {\rm eV^{2}},\quad \Delta m_{31}^{2}=2.511\times10^{-3}\, {\rm eV^{2}}.
\end{eqnarray}
The two Majorana phases $\alpha_{21}$ and $\alpha_{31}$ on the right hand side of Eq.~\eqref{eq:PMNS} are undetermined.
In our model, we only focus on the case with only one copy of $\mathbf{24}_{F}$, which contributes only two right-handed heavy leptons $N$ and $\Sigma$ for the generation of light neutrino masses. A well-known consequence is that, via the seesaw mechanism, the lightest neutrino is predicted to be massless, namely, $m_1 = 0$ in the normal ordering (NO, $m_{1}<m_{2}<m_{3}$) or $m_3=0$ in the inverted ordering (IO, $m_{3}<m_{1}<m_{2}$)  for neutrino masses. Once the lightest neutrino becomes massless, we are left with only one physical Majorana phase, which will be denoted as $\phi$, $\phi = (\alpha_{21}-\alpha_{31})/2$ for NO or $\phi = \alpha_{21}/2$ for IO.

We connect the Dirac Yukawa couplings $Y_{\rm I}$ and $Y_{\rm III}$ with the neutrino masses and the mixing angles. In the case with only one copy ${\bf 24}$-plet fermion, the light neutrino mass matrix is simplified into
\begin{eqnarray}
M_\nu = \frac{v^2}{2M_N} Y_{\rm I} Y_{\rm I}^T + \frac{v^2}{2M_\Sigma} Y_{\rm III} Y_{\rm III}^T\,,
\end{eqnarray}
where a global minus sign is dismissed. 
Keeping in mind Eq.~\eqref{eq:diagonalise_M_nu} and $U_{\rm PMNS} = U_\nu$ in the charged flavour diagonal basis, we apply Casas-Ibarra parametrization \cite{Casas:2001sr,Ibarra:2003up,Arhrib:2009xf} and obtain
\begin{eqnarray}
	Y_{\rm I}^{\alpha} &=& \frac{\sqrt{2M_{N}}}{v}(\sqrt{m_{2}}\,\cos z\,U_{\alpha 2}^{*}+\sqrt{m_{3}}\,\sin z\,U_{\alpha 3}^{*}) \,, \nonumber\\
	Y_{\rm III}^{\alpha} &=& \frac{\sqrt{2M_{\Sigma}}}{v}(-\sqrt{m_{2}}\,\sin z\,U_{\alpha 2}^{*}+\sqrt{m_{3}}\,\cos z\,U_{\alpha 3}^{*})
\end{eqnarray}
for the NO, where $U_{\alpha i}$ (for $\alpha =e, \mu, \tau$ and $i=1,2,3$) is the $(\alpha,i)$ entry of the PMNS matrix $U_{\rm PMNS}$, and $z = {\rm Re}z + i\, {\rm Im}z$ is a complex parameter.\\

\begin{figure}[t]
	\begin{center}
		\includegraphics[width=0.45\textwidth]{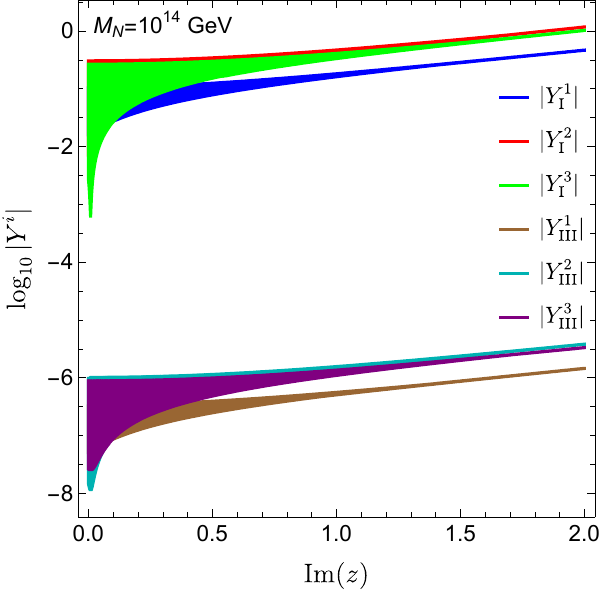}
		\includegraphics[width=0.45\textwidth]{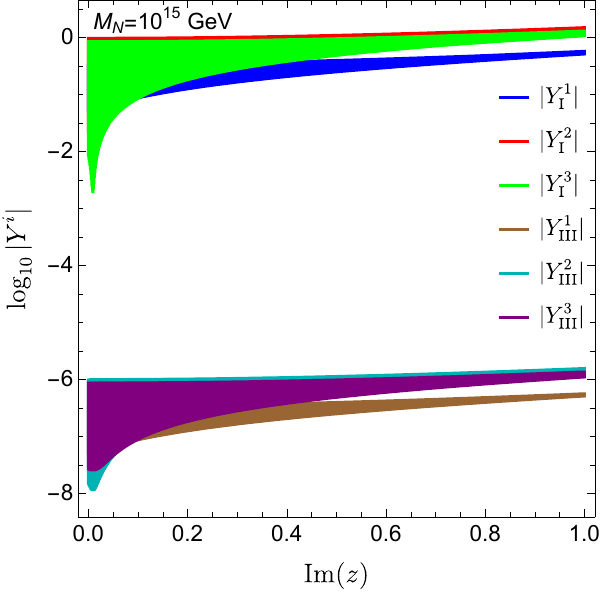}
		\caption{The absolute value of Yukawa couplings $|Y_{\rm I}^{i}|$ and $|Y_{\rm III}^{i}|$ as a function of Im(z) for NO, $M_{N}=10^{14}$ GeV (left panel) and $M_{N}=10^{15}$ GeV(right panel) are displayed, $\phi=0$ and $M_{\Sigma}=10^3~{\rm GeV}$ are determined.}
		\label{fig:flavor_mixing}
	\end{center}
\end{figure}

In the above analysis, we have set $M_{\Sigma}=10^3$ GeV and $M_{N}$ ranges in the allowed range consistent with gauge unification, i.e., ($3.20 \times 10^{13}$, $2.14 \times 10^{15}$ GeV).
Then, when ignoring the influence of the Majorana phase $\phi$, the Yukawa couplings $Y$ only depend on the complex parameter $z$. It is transparent that the real part of $z$ leads to a simple oscillatory nature, the imaginary part of $z$ actually affects the size of the couplings $Y$ most. In Fig.~\ref{fig:flavor_mixing}, we set $\phi=0$ and $M_{N}=10^{14},10^{15}$~GeV and show the relative size of the Yukawa couplings $|Y_{\rm I}^{i}|$ and $|Y_{\rm III}^{i}|$ as a function of ${\rm Im}(z)$ for NO with $M_{\Sigma}=10^3~{\rm GeV}$.
As we can see, $|Y_{\rm III}^{i}|$ is smaller than $|Y_{\rm I}^{i}|$ by a few orders of magnitude due to the large hierarchy between $M_{N}$ and $M_{\Sigma}$. When ${\rm Im}(z)$ is smaller then 2 (referring to $M_{N}=10^{14} {\rm GeV}$) or 1 (referring to $M_{N}=10^{15} {\rm GeV}$), $|Y_{\rm III}^{i}|$ is no more than $\mathcal{O}(1)$. In our case, the smallness of three left-handed neutrinos is attributed to the smallness of $|Y_{\rm III}^{i}|$ and the largeness of $M_{N}$.

\section{Proton decays}\label{sec:5}

The instability of the proton is one of the most attractive predictions of grand unified theories. In this part, our target is to understand if the model satisfies the current proton decay bounds and if it can be tested in future experiments. Partial decay widths of some typical channels of proton decay into a lepton and a meson are given by\cite{FileviezPerez:2004hn,FileviezPerez:2018dyf} 
\begin{eqnarray}
	\Gamma(p\rightarrow\pi^{0}e_{\alpha}^{+})&=& f_{\pi^0} \Big[ \big|\langle\pi^{0}|(ud)_{L}u_{L}|p\rangle \big|^{2} \big|c(e_{\alpha}^{c},d)\big|^2+ 
	\big|\langle\pi^{0}|(ud)_{R}u_{L}|p\rangle \big|^{2} \big| c(e_{\alpha},d^{c})\big|^2 \Big] \,, \nonumber\\
\Gamma(p\rightarrow K^{0}e_{\alpha}^{+}) &=& f_{K^0} \big|\langle K^{0}|(us)_{R}u_{L}|p\rangle \big|^{2} \Big[ \big| c(e_{\alpha},s^{c}) \big|^{2}+\big| c(e_{\alpha}^{c},s)\big|^{2}\Big] \,, \nonumber\\
\Gamma(p\rightarrow K^{+}\bar{\nu})&=& f_{K^+} \sum_{i} \Big[ \big| \langle K^{+}|(us)_{R}d_{L}|p\rangle \big|^{2} \big| c(\nu_{i},d,s^{c}) \big|^2
+ \big| \langle K^{+}|(ud)_{R}s_{L}|p\rangle \big|^{2} \big| c(\nu_{i},s,d^{c}) \big|^2 \Big] \,, \nonumber\\
	\Gamma(p\rightarrow \pi^{+}\bar{\nu})&=& f_{\pi^+} \sum_{i} \big| \langle \pi^{+}|(du)_{R}d_{L}|p\rangle \big|^{2} \big| c(\nu_{i},d,d^{c}) \big|^2 \,.
\end{eqnarray}
Here $f_M $ is the phase space contribution ignoring charged lepton masses,
\begin{eqnarray}
f_M = \frac{1}{8 \pi} \left(1-\frac{m_{M}^{2}}{m_{p}^{2}}\right)^2 \left(\frac{g_{\rm GUT}}{\sqrt{2}M_{\rm GUT}}\right)^4 A_{\rm LD}^2A_{\rm SD}^2 \,. 
\end{eqnarray}
$A_{\rm LD}$ and $A_{\rm SD}$ parametrising the long-distance and short-distance effect of the baryon-number-violating operators \cite{Nath:2006ut}, accounting for renormalisation contribution of the QCD running from $m_{t}$ to the proton mass scale
\begin{eqnarray}
	A_{\rm LD}=\left(\frac{\alpha_{3}(m_{p})}{\alpha_{3}(m_{c})}\right)^{\frac{2}{9}}\left(\frac{\alpha_{3}(m_{c})}{\alpha_{3}(m_{b})}\right)^{\frac{6}{25}}\left(\frac{\alpha_{3}(m_{b})}{\alpha_{3}(m_{t})}\right)^{\frac{6}{23}}
\end{eqnarray}
and that from the GUT scale down to $m_{t}$
\begin{eqnarray}
	A_{\rm SD}&=&\prod_{I=1,2,...}  \left(\frac{\alpha_{1}(M_{I})}{\alpha_{1}(M_{I-1})}\right)^{-\frac{23}{30(a_{1}^{\rm SM}+\Delta a_{1}^I)}}\left(\frac{\alpha_{2}(M_{I})}{\alpha_{2}(M_{I-1})}\right)^{-\frac{3}{2(a_{2}^{\rm SM}+\Delta a_{2}^I)}}
	\left(\frac{\alpha_{3}(M_{I})}{\alpha_{3}(M_{I-1})}\right)^{-\frac{4}{3(a_{3}^{\rm SM}+\Delta a_{3}^I)}} \,,
\end{eqnarray}
respectively, where we considered the extra matter at a single intermediate scale $M_{I}$ and $\Delta a_{i}\equiv \sum_{I}a_{i}^{I}$.\\
The revelant hadronic matrix elements can be obtained by using QCD lattice simulation. We use the best-fit values reported in the recent paper\cite{Aoki:2017puj}, i.e., 
\begin{eqnarray}
	\langle\pi^{0}|(ud)_Lu_L|p\rangle=0.134(5)(16) \text{ GeV}^2,\qquad&& \langle\pi^{0}|(ud)_Ru_L|p\rangle=-0.131(4)(13) \text{ GeV}^2,\nonumber\\
	\langle\pi^{+}|(ud)_Rd_L|p\rangle=-0.186(6)(18) \text{ GeV}^2,\qquad&& \langle\pi^{+}|(ud)_Ld_L|p\rangle=-0.189(6)(22) \text{ GeV}^2,\nonumber\\
	\langle K^{0}|(us)_Ru_L|p\rangle=0.103(3)(11) \text{ GeV}^2,\qquad&&
	\langle K^{0}|(us)_Lu_L|p\rangle=0.057(2)(6) \text{ GeV}^2,\nonumber\\
	\langle K^{+}|(us)_Rd_L|p\rangle=-0.049(2)(5) \text{ GeV}^2,\qquad&&
	\langle K^{+}|(us)_Ld_L|p\rangle=0.041(2)(5) \text{ GeV}^2,\nonumber\\
	\langle K^{+}|(ud)_Rs_L|p\rangle=-0.134(4)(14) \text{ GeV}^2,\qquad&&
	\langle K^{+}|(ud)_Ls_L|p\rangle=0.139(4)(15) \text{ GeV}^2,\nonumber\\
	\langle K^{+}|(ds)_Ru_L|p\rangle=-0.054(2)(6) \text{ GeV}^2,\qquad&&
	\langle K^{+}|(ds)_Lu_L|p\rangle=-0.098(3)(10) \text{ GeV}^2,\nonumber\\
	\langle \eta|(ud)_Ru_L|p\rangle=0.006(2)(3) \text{ GeV}^2,\qquad&&
	\langle \eta|(ud)_Lu_L|p\rangle=0.113(3)(12) \text{ GeV}^2.
\end{eqnarray}
The $c$ coefficients appearing in these formulas are \cite{FileviezPerez:2004hn}
\begin{eqnarray}
	&&c(e_{\alpha}^{c},d)=(U_{u}^{\prime T}U_{u})_{11}(U_{d}^{T}U_{e}^{\prime})_{1\alpha}+(U_{u}^{\prime T}U_{d})_{11}(U_{u}^{T}U_{e}^{\prime})_{1\alpha}\,, \nonumber\\
	&&c(e_{\alpha}^{c},s)=(U_{u}^{\prime T}U_{u})_{11}(U_{d}^{T}U_{e}^{\prime})_{2\alpha}+(U_{u}^{\prime T}U_{d})_{12}(U_{u}^{T}U_{e}^{\prime})_{1\alpha}\,, \nonumber\\
	&&c(e_{\alpha},d^{c})=(U_{u}^{\prime T}U_{u})_{11}(U_{d}^{\prime T}U_{e})_{1\alpha}\,, \nonumber\\
	&&c(e_{\alpha},s^{c})=(U_{u}^{\prime T}U_{u})_{11}(U_{d}^{\prime T}U_{e})_{2\alpha}\,, \nonumber\\
	&&c(\nu_{i},d,d^{c})=(U_{u}^{\prime T}U_{d})_{11}(U_{d}^{\prime T}U_{\nu})_{1i}\,, \nonumber\\
	&&c(\nu_{i},d,s^{c})=(U_{u}^{\prime T}U_{d})_{11}(U_{d}^{\prime T}U_{\nu})_{2i}\,, \nonumber\\
	&&c(\nu_{i},s,d^{c})=(U_{u}^{\prime T}U_{d})_{12}(U_{d}^{\prime T}U_{\nu})_{1i}\,,
\end{eqnarray}
where $\alpha,\beta=1,2$.
There are a large number of free parameters introduced by the unitary matrices.
In the above section, we have used a basis where $U_e = U_e' = {\bf 1}$. Even in this case, we still have large redundancies. 
In the following phenomenological discussion, we will further concentrate on the following three specified scenarios.
\begin{itemize}
\item[S1)] Diagonal down-type quark Yukawa couplings, $Y_{d} = \hat{Y_{d}}$, leading to  $U_d = U_d' = {\bf 1}$. 
\item[S2)] Diagonal up-type quark Yukawa couplings, $Y_{u} = \hat{Y_{u}}$,  leading to $U_u = U_u' = {\bf 1}$.
\item[S3)] Hermitian quark Yukawa couplings, $Y_d = Y_d^\dag$ and $Y_u = Y_u^\dag$, leading to $U_u = U_u'$ and $U_d = U_d'$.
\end{itemize}
Detailed discussions on these scenarios are given below.

\subsection*{S1) $Y_d = \hat{Y}_d$}
In this scenario, we have $U_{u} = U_d V_{\rm CKM}^\dag = V_{\rm CKM}^{\dag}$. $U_{u}^{\prime}$ is in general different from $U_{u}$. Sizes of each entries of $V_{\rm CKM}$ satisfy $|V_{cd}|\simeq|V_{us}|\equiv\lambda,|V_{ts}|\simeq|V_{cb}|\sim\lambda^{2},|V_{td}|\sim\lambda^{3}$ and $|V_{ub}|\sim|V_{td}|/2$, where $\lambda = \sin \theta_C$ and $\theta_C$ is the Cabibbo angle. When calculating the $c$ coefficients in proton decay lifetime, we consider precision up to the order $\mathcal{O}(\lambda)$ and ignore smaller quantities. Then, proton partial decay widths can be expressed more straightforwardly,
\begin{eqnarray} \label{eq:pd_S1}
\Gamma(p\rightarrow\pi^{0}e^{+}) &\approx& f_{\pi^0}
\Big[\big|\langle\pi^{0}|(ud)_{L}u_{L}|p\rangle\big|^{2}
\big|2(U_{u}^\prime)_{11}V_{ud}^*+(U_{u}^\prime)_{21}V_{us}^*\big|^{2} \nonumber\\
&&+\big|\langle\pi^{0}|(ud)_{R}u_{L}|p\rangle \big|^{2}
\big|(U_{u}^\prime)_{11}V_{ud}^*+(U_{u}^\prime)_{21}V_{us}^*\big|^2 \Big] \,,\nonumber\\
\Gamma(p\rightarrow \pi^{0}\mu^{+}) &\approx& f_{\pi^0} \big|\langle \pi^{0}|(ud)_{L}u_{L}|p\rangle \big|^{2} \big|(U_u^\prime)_{11}V_{us}^*\big|^{2} \,, \nonumber\\
\Gamma(p\rightarrow K^{0}e^{+}) &\approx& f_{K^0} \big|\langle K^{0}|(us)_{R}u_{L}|p\rangle \big|^{2} \big|(U_u^\prime)_{21}V_{ud}^*\big|^{2} \,, \nonumber\\
\Gamma(p\rightarrow K^{0}\mu^{+})&\approx& f_{K^0}
\big|\langle K^{0}|(us)_{R}u_{L}|p\rangle \big|^{2} \Big[|(U_{u}^\prime)_{11}V_{ud}^*+(U_{u}^\prime)_{21}V_{us}^*|^{2}+|(U_{u}^\prime)_{11}V_{ud}^*+2(U_{u}^\prime)_{21}V_{us}^*|^{2}\Big] \,, \nonumber\\
\Gamma(p\rightarrow \pi^{+}\bar{\nu}) &\approx& f_{\pi^+} \big|\langle \pi^{+}|(du)_{R}d_{L}|p\rangle \big|^{2} \big|(U_u^\prime)_{11}\big|^{2} \,, \nonumber\\
\Gamma(p\rightarrow K^{+}\bar{\nu}) &\approx& f_{K^+} \Big[\big| \langle K^{+}|(us)_{R}d_{L}|p\rangle \big|^{2} \big|(U_u^\prime)_{11} \big|^{2}+
\big| \langle K^{+}|(ud)_{R}s_{L}|p\rangle\big|^{2} \big|(U_u^\prime)_{21} \big|^{2}\Big]\,.
\label{proton_decay_S1}
\end{eqnarray}
Obviously, we are left with only two free parameters $(U_{u}^\prime)_{11}$ and $(U_{u}^\prime)_{21}$, which are relevant to some degrees of freedom associated with the right-handed up-type quark fields. To see more clearly the dependence of proton lifetime upon the parameters, we assume $M_{\Sigma}=10^3~{\rm GeV}$ and $M_{\rm GUT}=2 \times 10^{15}~{\rm GeV}$, values of which are consistent with gauge unification discussed in the last section. The proton partial decay lifetime can be obtained numerically as
\begin{eqnarray}	  
\tau(p\rightarrow\pi^{0}e^{+}) &\approx& \frac{7.1 \times 10^{32} ~{\rm yr}}{|(U_u^\prime)_{11}^2+0.2747(U_u^\prime)_{11}(U_u^\prime)_{21}+0.0209(U_u^\prime)_{21}^2|} \,,\nonumber\\
\tau(p\rightarrow\pi^{0}\mu^{+}) &\approx& \frac{6.6 \times 10^{34} ~{\rm yr}}{|(U_u^\prime)_{11}^2|} \,,\nonumber\\
\tau(p\rightarrow K^{0}e^{+}) &\approx& \frac{1.1 \times 10^{34}~{\rm yr}}{|(U_u^\prime)_{21}^2|} \,,\nonumber\\
\tau(p\rightarrow K^{0}\mu^{+}) &\approx& \frac{5.7 \times 10^{33}~{\rm yr}}{|(U_u^\prime)_{11}^2+0.6911(U_u^\prime)_{11}(U_u^\prime)_{21}+0.1327(U_u^\prime)_{21}^2|} \,, \nonumber\\
\tau(p\rightarrow \pi^{+}\bar{\nu}) &\approx& \frac{1.7 \times 10^{33}~{\rm yr}}{|(U_u^\prime)_{11}^2|} \,,\nonumber\\
\tau(p\rightarrow K^{+}\bar{\nu}) &\approx& \frac{4.7 \times 10^{34}~{\rm yr}}{|(U_u^\prime)_{11}^2+7.4622(U_u^\prime)_{21}^2|} \,,
\label{proton_decay_S1_num}
\end{eqnarray}
where we have used values of CKM matrix elements $V_{ud},V_{us}$ from Particle Data Group \cite{Workman:2022ynf}. A short lifetime for $p\rightarrow\pi^{0}e^{+}$ is predicted if $U_{u}' \simeq 1$. Namely, to generate a signal consistent with current data, the unitary matrix $U_{u}'$ must deviates from the diagonal matrix. 

\begin{figure}[t!]
	\begin{center}
		\includegraphics[width=0.48\textwidth]{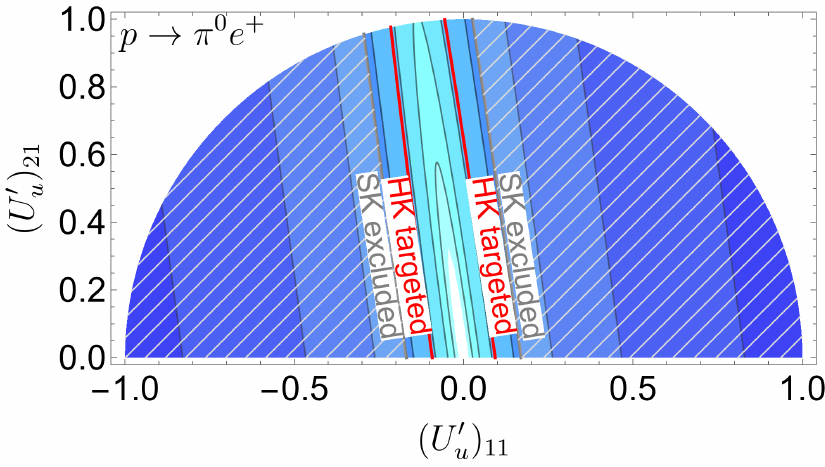}
		\includegraphics[width=0.48\textwidth]{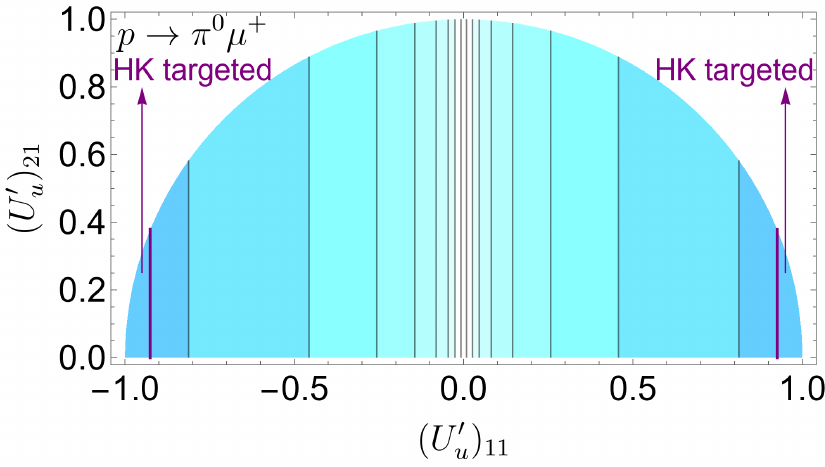}\\[3mm]
		\includegraphics[width=0.48\textwidth]{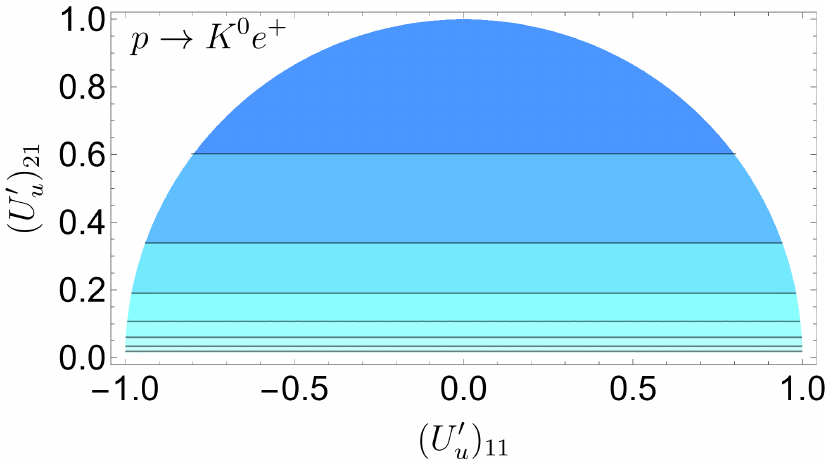}
		\includegraphics[width=0.48\textwidth]{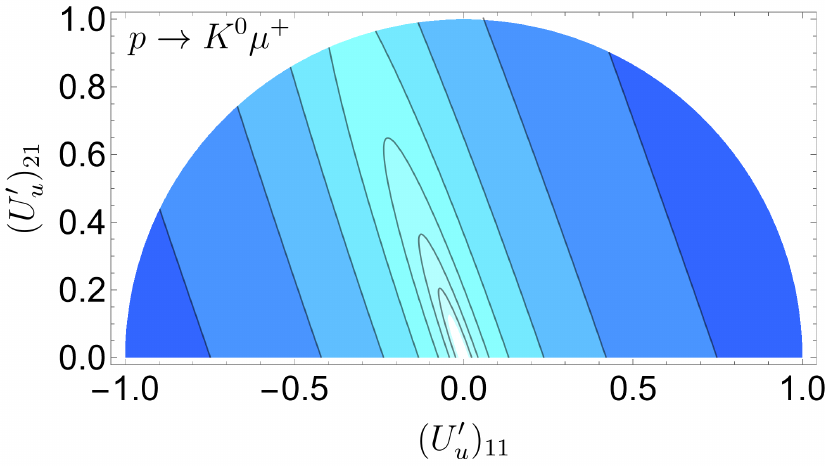}\\[3mm]
		\includegraphics[width=0.48\textwidth]{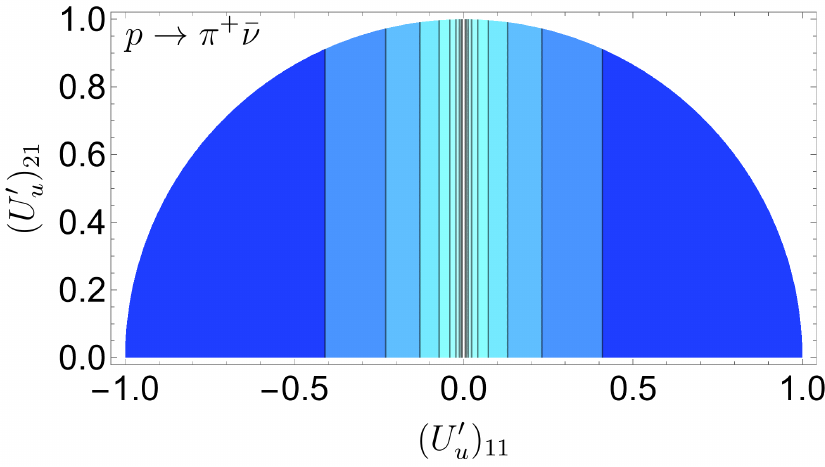}
		\includegraphics[width=0.48\textwidth]{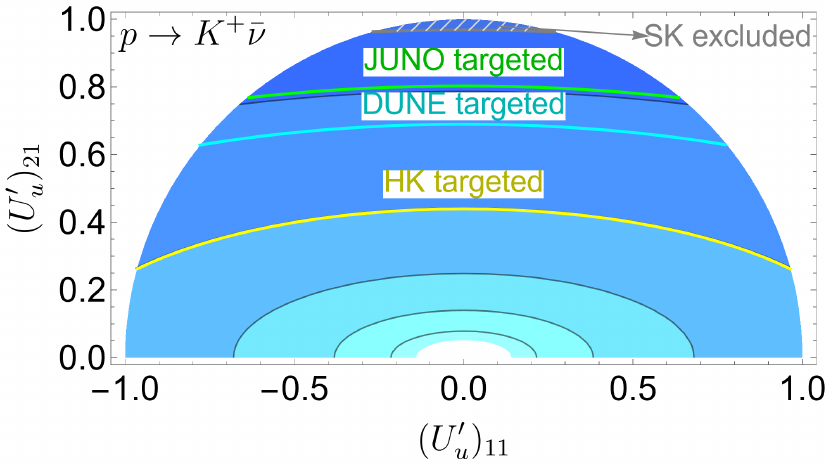}
		\includegraphics[width=0.4\textwidth]{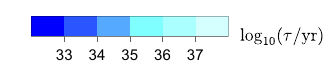}
		\caption{Predictions for the partial lifetime of proton decay in different channels $p\to\pi^{0}e^{+}$, $\pi^{0}\mu^{+}$, $K^{0}e^{+}$, $K^{0}\mu^{+}$, $\pi^{+}\bar{\nu}$, and $K^{+}\bar{\nu}$. $M_{\rm GUT}=2 \times 10^{15}~{\rm GeV}$ and $M_{\Sigma}=10^3 {\rm GeV}$ are assumed. The region hatched in gray curves are excluded by SK, referring to $\tau(p\rightarrow\pi^{0}e^{+})>2.4\times 10^{34}$ years (top-left) \cite{Super-Kamiokande:2020wjk}, $\tau(p\rightarrow K^{+}\bar{\nu})>6.6\times 10^{33}$ years (botton-right) \cite{Takhistov:2016eqm}. SK also set the lower bounds for other channels, $\tau(p\rightarrow K^{0}\mu^{+})>3.6\times 10^{33}$ years\cite{Super-Kamiokande:2022egr}, $\tau(p\rightarrow \pi^{+}\bar{\nu})>3.9\times 10^{32}$ years \cite{Super-Kamiokande:2013rwg}, $\tau(p\rightarrow\pi^{0}\mu^{+})>1.6\times 10^{34}$ years \cite{Super-Kamiokande:2020wjk} and $\tau(p\rightarrow K^{0}\mu^{+})>3.6\times 10^{33}$ years \cite{Super-Kamiokande:2022egr}, which are consistent with all the flavour space. The green curve shows future sensitivity in JUNO, $\tau(p\rightarrow K^{+}\bar{\nu})>9.6\times 10^{33}$ years\cite{JUNO:2022qgr}. The cyan curve shows that in DUNE, i.e., $\tau(p\rightarrow K^{+}\bar{\nu})>1.3\times 10^{34}$ years\cite{DUNE:2020ypp}. The red, yellow and purple curves show the sensitivities for different channels in HK, i.e., $\tau(p\rightarrow\pi^{0}e^{+})>8\times 10^{34}$ years\cite{Yokoyama:2017mnt}, $\tau(p\rightarrow K^{+}\bar{\nu})>3.2\times 10^{34}$ years\cite{Hyper-Kamiokande:2018ofw}, $\tau(p\rightarrow\pi^{0}\mu^{+})>7.7\times 10^{34}$ years\cite{Hyper-Kamiokande:2018ofw}.}
		\label{fig:S1_v1}
	\end{center}
\end{figure}

We perform a numerical scan for illustration. $(U_u^\prime)_{11}$ and $(U_u^\prime)_{21}$ are assumed to be real and $(U_u^\prime)_{11}  \in (-1,1)$ and $(U_u^\prime)_{21} \in (0,1)$ are restricted. The unitarity of $U_u^\prime$ requires $|(U_u^\prime)_{11}|^2+|(U_u^\prime)_{21}|^2\leq 1$. An explicit analysis should include a relative phase between $(U_u^\prime)_{11}$ and $(U_u^\prime)_{21}$ if they are relaxed to be complex. This phase is less important and ignored here. Instead, the positive and negative signs for $(U_u^\prime)_{11}$ refer to two extremal cases for the relative phase at 0 and $\pi$, respectively. Obviously the proton lifetime tends to infinity as $(U_u^\prime)_{11},(U_u^\prime)_{21}$ tend to zero. Then we show the predictions for the six channels above in Fig.~\ref{fig:S1_v1}. 
Here, we observe that for the $K^{+}\bar{\nu}$ channel, a large range of the parameter space of $(U_u^\prime)_{11}$ and $(U_u^\prime)_{21}$ is still consistent with the Hyper-Kamiokande (HK) bound. However, most of this range has been excluded in the $\pi^0 e^+$ channel measurement, as expected from the analytical formula in Eq.~\eqref{eq:pd_S1}. The prediction of proton decay lifetime for the channel $p\rightarrow K^{0}\mu^{+}$ is much longer than the upper bound set by Super-Kamiokande (SK). As we can see, the prediction of proton decay lifetime for the channel $p\rightarrow \pi^{+}\bar{\nu}$ is bigger than the limit of SK completely. 

\begin{figure}[t]
	\begin{center}
		\includegraphics[width=0.45\textwidth]{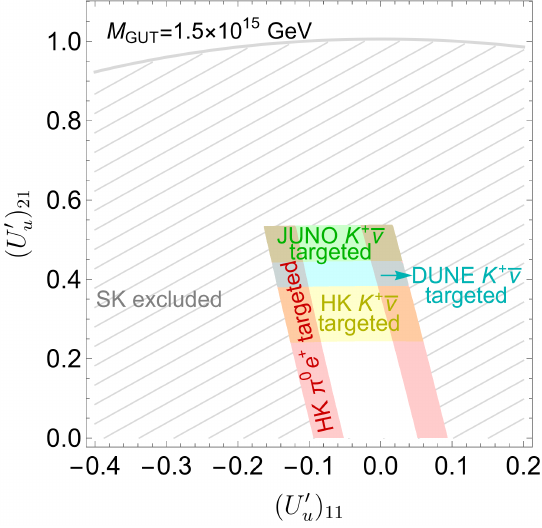}
		\includegraphics[width=0.45\textwidth]{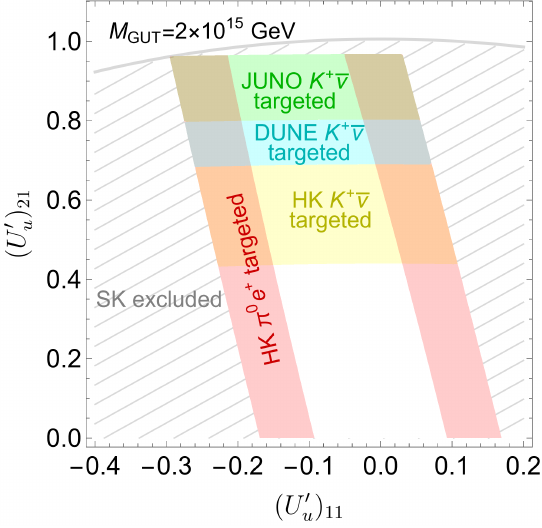}
		\caption{Correlation between $(U_u^\prime)_{11}$ and $(U_u^\prime)_{21}$ within future experiments scope.}
		\label{Uu}
	\end{center}
	\label{fig:S1_v2}
\end{figure}

We further check the testability of the the flavour space in light of the JUNO, DUNE and HK experiments. We zoom in Fig.~\ref{Uu}  the parameter space of $(U_u^\prime)_{11}$ and $(U_u^\prime)_{21}$ within the touch of these experiments. The parameter space  targeted by HK via the  $\pi^{0}e^{+}$ channel is intersected with that targeted by JUNO, DUNE and HK via the $K^+\bar{\nu}$ channel, but each  meaurements all have their own respectively targeted regions. It is notably to menion that as $M_{\rm GUT}$ gets larger, the allowed parameter space increases, $\tau(p\rightarrow K^+\bar{\nu})$ gets larger and exceeds JUNO's ability, even exceeds DUNE's ability. To sum up, there is hope to test this theory in future proton decay experiments, especially in the JUNO, DUNE and HK experiments.

\subsection*{S2) $Y_{u} = \hat{Y_{u}}$}

In S2), we have $U_{d} = U_u V_{\rm CKM} = V_{\rm CKM}$. $U_{d}^{\prime}$ in general should be different from $U_{d}$. Similar to S1), proton partial decay widths can be written in simple way, e.g.,
\begin{eqnarray}
	\Gamma(p\rightarrow\pi^{0}e^{+}) &\approx& f_{\pi^0} \Big[ \big|\langle\pi^{0}|(ud)_{L}u_{L}|p\rangle \big|^{2} 4 \big| V_{ud} \big|^{2} + \big|\langle\pi^{0}|(ud)_{R}u_{L}|p\rangle \big|^{2} \big|(U_{d}^{\prime})_{11} \big|^{2} \Big]\,, \nonumber\\
	\Gamma(p\rightarrow K^{+}\bar{\nu}) &\approx& f_{K^+} \Big[ \big| \langle K^{+}|(us)_{R}d_{L}|p\rangle \big|^{2} \big|V_{ud}\big|^{2} +
	\big|\langle K^{+}|(ud)_{R}s_{L}|p\rangle \big|^{2} \big| V_{us} \big|^{2}  \Big] \,.
\end{eqnarray}
The numerical expression of the above equation is
\begin{eqnarray}
\tau(p\rightarrow\pi^{0}e^{+})&\approx&\frac{3.4 \times 10^{33}~{\rm yr}}{|(U_d^\prime)_{11}^2+3.9737|}\,,\nonumber\\
\tau(p\rightarrow K^{+}\bar{\nu}) &\approx& 3.58\times 10^{34}~{\rm yr}\,.
\end{eqnarray}
once $M_\Sigma = 10^3$~GeV and $M_{\rm GUT} = 2\times 10^{15}$~GeV are fixed. Here, $\Gamma(p\rightarrow K^{+}\bar{\nu})$ is just a number which is not relevant to $U_d^\prime$. However, we have checked that the prediction for the $p\rightarrow\pi^{0}e^{+}$ channel is always less than $8.7 \times 10^{32}$~years, which is obviously not meeting current limits of SK, $\tau(p\rightarrow\pi^{0}e^{+})>2.4\times 10^{34}$ years \cite{Super-Kamiokande:2020wjk}. Therefore, this scenario is excluded.

\subsection*{S3)  $Y_d^\dag = Y_d$ and $Y_u^\dag = Y_u$}

\begin{figure}[t]
	\begin{center}
		\includegraphics[width=0.48\textwidth]{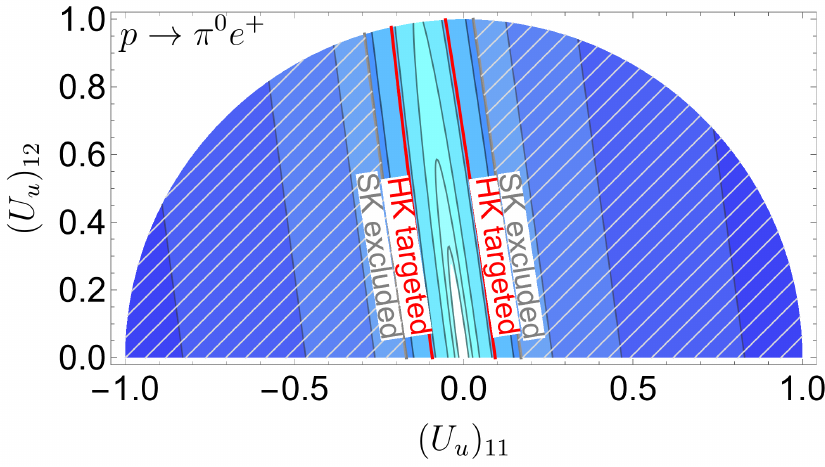}
		\includegraphics[width=0.48\textwidth]{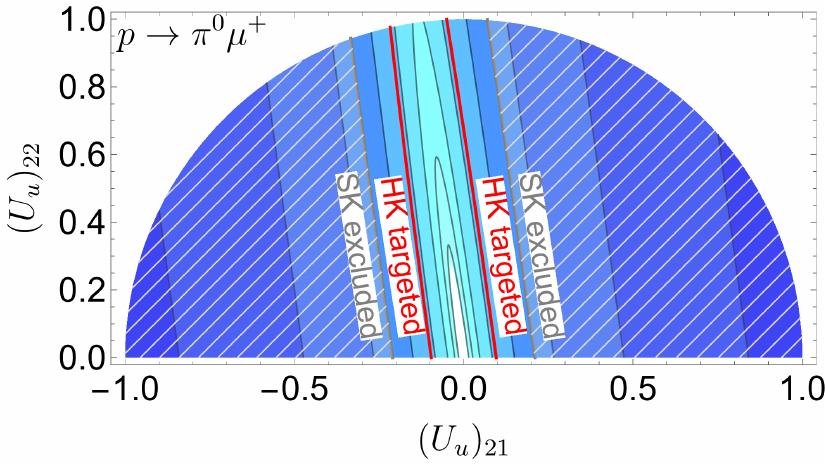}\\[3mm]
		\includegraphics[width=0.48\textwidth]{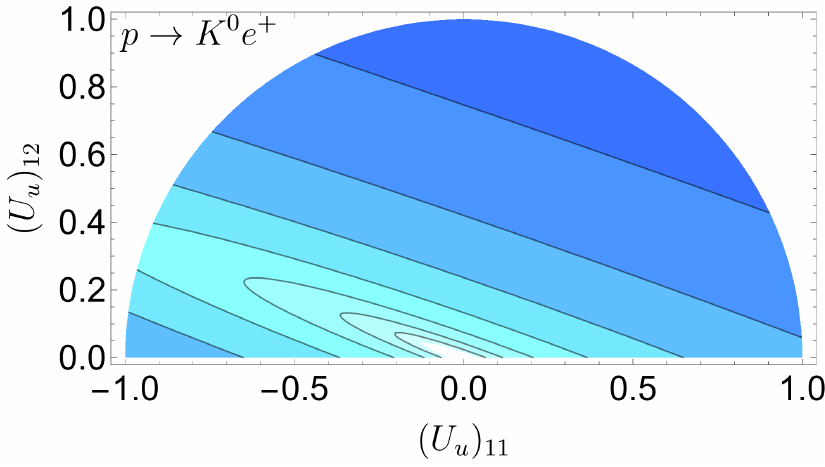}
		\includegraphics[width=0.48\textwidth]{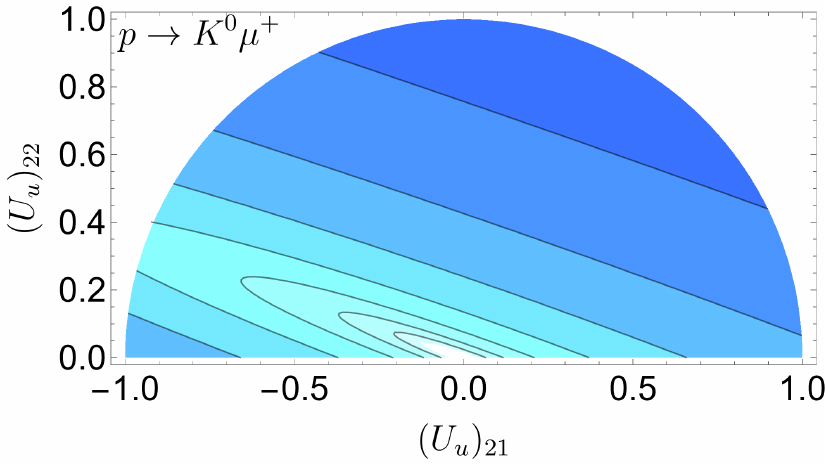}
		\includegraphics[width=0.4\textwidth]{fig/plegend.pdf}
		\caption{Predictions for the proton decay in S3). $M_{\rm GUT}=2 \times 10^{15}~{\rm GeV}$ and $M_{\Sigma}=10^3~{\rm GeV}$ are assumed. The region hatched in gray curves are excluded by SK, i.e., $\tau(p\rightarrow\pi^{0}e^{+})>2.4\times 10^{34}$ years \cite{Super-Kamiokande:2020wjk} (top-left) and $\tau(p\rightarrow\pi^{0}\mu^{+})>1.6\times 10^{34}$ years \cite{Super-Kamiokande:2020wjk} (top-right). Future sensitivities in HK are shown in red curves, i.e., $\tau(p\rightarrow\pi^{0}e^{+})>8\times 10^{34}$ years\cite{Yokoyama:2017mnt}, $\tau(p\rightarrow\pi^{0}\mu^{+})>7.7\times 10^{34}$ years\cite{Hyper-Kamiokande:2018ofw}. In the bottom panels, current experimental lower bounds for $p\rightarrow K^0 e^+$ and $K^0 \mu^+$, i.e., $\tau(p\rightarrow K^{0}e^{+})>1\times 10^{33}$ years \cite{Super-Kamiokande:2005lev} and $\tau(p\rightarrow K^{0}\mu^{+})>3.6\times 10^{33}$ years \cite{Super-Kamiokande:2022egr}, are smaller than the prediction for all parameter space of $U_u$ and thus shows no constraints to the latter. The partial lifetime for $p \to \pi^+ \bar{\nu}$ and $K^+ \bar{\nu}$ is independent of $U_u$ and thus not shown.
}
	\label{fig:S3}
	\end{center}
\end{figure}

In S3), we have $U_{d} = U_u V_{\rm CKM}$. $U_u$ is a free unitary matrix. Similar to the discussion in S1), we obtain proton partial decay widths here as
\begin{eqnarray}
	\Gamma(p\rightarrow\pi^{0}e^{+}) &\approx& f_{\pi^0} \Big[
	\big|\langle\pi^{0}|(ud)_{L}u_{L}|p\rangle \big|^{2} \big| 2(U_u)_{11}V_{ud}+(U_u)_{12}V_{cd} \big|^{2} \nonumber\\
	&&+ \big| \langle\pi^{0}|(ud)_{R}u_{L}|p\rangle \big|^{2} \big| (U_u)_{11}V_{ud}+(U_u)_{12}V_{cd} \big|^{2} \Big] \,,\nonumber\\
	\Gamma(p\rightarrow\pi^{0}\mu^{+}) &\approx& f_{\pi^0} \Big[
	\big|\langle\pi^{0}|(ud)_{L}u_{L}|p\rangle \big|^{2} \big| 2(U_u)_{21}V_{ud}+(U_u)_{22}V_{cd} \big|^{2} \nonumber\\
	&&+ \big| \langle\pi^{0}|(ud)_{R}u_{L}|p\rangle \big|^{2} \big| (U_u)_{21}V_{ud}+(U_u)_{22}V_{cd} \big|^{2} \Big] \,,\nonumber\\
	\Gamma(p\rightarrow K^{0}e^{+}) &\approx& f_{K^0} \big| \langle K^{0}|(us)_{R}u_{L}|p\rangle \big|^{2} 
	\Big[ \big| 2(U_u)_{11}V_{us}+(U_u)_{12}V_{cs} \big|^{2} + \big| (U_u)_{11}V_{us}+(U_u)_{12}V_{cs} \big|^{2} \Big] \,,\nonumber\\
	\Gamma(p\rightarrow K^{0}\mu^{+}) &\approx& f_{K^0} \big| \langle K^{0}|(us)_{R}u_{L}|p\rangle \big|^{2} 
	\Big[ \big| 2(U_u)_{21}V_{us}+(U_u)_{22}V_{cs} \big|^{2} + \big| (U_u)_{21}V_{us}+(U_u)_{22}V_{cs} \big|^{2} \Big] \,,\nonumber\\
	\Gamma(p\rightarrow\pi^{+}\bar{\nu}) &\approx& f_{\pi^+} \big|\langle\pi^{+}|(du)_{R}d_{L}|p\rangle \big|^{2} \big| V_{ud} \big|^{2} \,,\nonumber\\
	\Gamma(p\rightarrow K^{+}\bar{\nu}) &\approx& f_{K^+} \Big[ \big| \langle K^{+}|(us)_{R}d_{L}|p\rangle \big|^{2} \big| V_{ud} \big|^{2} + \big| \langle K^{+}|(ud)_{R}s_{L}|p\rangle \big|^{2} \big| V_{us} \big|^{2} \Big] \,, 
\end{eqnarray}
In this scenario, four free parameters are involved, $(U_{u})_{11}$, $(U_{u})_{12}$, $(U_{u})_{21}$ and $(U_{u})_{22}$. Formulas for $p \to \pi^0 e^+$ and $p \to \pi^0 \mu^+$, as well as those for $p \to K^0 e^+$ and $p \to K^0 \mu^+$, are almost the same except the replacements $(U_u)_{11} \to (U_u)_{21}$ and $(U_u)_{12} \to (U_u)_{22}$. 
$\Gamma(p\rightarrow K^{+}\bar{\nu})$ and $\Gamma(p\rightarrow \pi^{+}\bar{\nu})$ are independent of any entries of $U_{u}$. By fixing $M_{\Sigma}=10^3~{\rm GeV}$ and $M_{\rm GUT}=2\times 10^{15}~{\rm GeV}$, we obtain 
\begin{eqnarray}	  
\tau(p\rightarrow\pi^{0}e^{+}) &\approx& \frac{7.1 \times 10^{32}~{\rm yr}}{|(U_u)_{11}^2+0.2707(U_u)_{11}(U_u)_{12}+0.0203(U_u)_{12}^2|} \,,\nonumber\\
\tau(p\rightarrow K^{0}e^{+}) &\approx& \frac{4.3 \times 10^{34}~{\rm yr}}{|(U_u)_{11}^2+5.2162(U_u)_{11}(U_u)_{12}+7.5581(U_u)_{12}^2|} \,,\nonumber\\
\tau(p\rightarrow\pi^{0}\mu^{+}) &\approx& \frac{7.1 \times 10^{32}~{\rm yr}}{|(U_u)_{21}^2+0.2707(U_u)_{21}(U_u)_{22}+0.0203(U_u)_{22}^2|} \,,\nonumber\\
\tau(p\rightarrow K^{0}\mu^{+}) &\approx& \frac{4.3 \times 10^{34}~{\rm yr}}{|(U_u)_{21}^2+5.2162(U_u)_{21}(U_u)_{22}+7.5581(U_u)_{22}^2|} \,,\nonumber\\
\tau(p\rightarrow \pi^{+}\bar{\nu}) &\approx& 1.8 \times10^{33}~{\rm yr} \,,\nonumber\\
\tau(p\rightarrow K^{+}\bar{\nu}) &\approx& 3.58 \times 10^{34}~{\rm yr} \,,
\end{eqnarray}
where values of CKM matrix elements $V_{ud},V_{cd},V_{us},V_{cs}$ have been taken from Particle Data Group(PDG)\cite{Workman:2022ynf}. We show the dependence of proton partial lifetime upon entries of $U_u$ in Fig.~\ref{fig:S3}. 
The last two channels, which are independent of $U_u$, predicts fixed partial lifetime consistent with current SK bounds \cite{Super-Kamiokande:2013rwg,Takhistov:2016eqm}. In particular, the $p \to K^+ \bar{\nu}$ channel is beyond the future sensitivities of HK \cite{Hyper-Kamiokande:2018ofw}. 

To end this section, we comment on the case with 2HDM. Following the discussion in the end of Section~\ref{sec:3}, the main effect of including one more Higgs doublet at the electroweak scale is an enhancement of the GUT scale, but not to the flavour space. We expect that, once the GUT scale fixed at the same scale, the restriction on the parameter space of $(U_{u}')_{11}$ and $(U_{u}')_{21}$ in Figs.~\eqref{fig:S1_v1} and \eqref{fig:S1_v2}, as well as that of $(U_{u})_{11}$ and $(U_{u})_{21}$ in Fig.~\eqref{fig:S3}, keep almost the same. Assuming a higher GUT scale, e.g., $M_{\rm GUT} = 3\times 10^{15}~{\rm GeV}$, which are not consistent with SM but 2HDM, a longer proton decay is predicted and the restriction on the flavour space is weaker.

\section{Conclusion}

We discussed economical extensions of $SU(5)$ model by including extra ${\bf 24}_F$ fermion multiplets. These multiplets include electroweak singlets and triplets, which may play important roles in light neutrino mass generation through type-I or type-III seesaw mechanisms. Including the ${\bf 24}_F$ fermions saves the gauge unification and predicts proton decay lifetime compatible with current experimental bounds.  We focus on the minimal case that only one copy of ${\bf 24}_F$ is introduced. Light neutrino masses are generated via type-(I+III) seesaw and the lightest neutrino keeps massless. The electroweak singlet $N$ and triplet fermion $\Sigma$ are restricted to be around the canonical seesaw scale and TeV scale, respectively. We have checked that all data of lepton flavour mixing can be reproduced, but large hierarchy between Yukawa for $N$ and that for $\Sigma$ has to be included. {The mass splitting for ${\bf 24}_F$ is achieved via the Higgs in the adjoint representation, ${\bf 24}_{\Phi}$, which was originally introduced in the George-Glashow model for the breaking of $SU(5)$.}
In the minimal case,  these Higgs are assumed to be heavy enough and do not contribute to RG running below the GUT scale. {Including the contribution of only one copy ${\bf 24}_F$ in the RG running is enough to achieve a sufficiently high GUT scale.} 

We further study contribution of free parameters in the flavour space to different channels of nucleon decays. We have considered three scenarios: S1) down-type quark Yukawa matrix is diagonal; S2) up-quark Yukawa matrix is diagonal; and S3) both Yukawa matrices are Hermitian. In S1), only two free parameters in the flavour space contribute to the proton decay channels we discussed in this work, $p \to \pi^0 e^+, \pi^0 \mu^+, K^0 e^+, K^0 \mu^+, \pi^+ \bar{\nu}, K^+ \bar{\nu}$.
Combining $\pi^{0}e^{+}$ and $K^{+}\overline{\nu}$ channel for analysis excludes most of the parameter space of S1), but there is still some space for future experiments to test, such as JUNO and HK. S2) predicts a short partial lifetime for $p \to \pi^0 e^+$ and  has fully been excluded by SK. In S3), more free parameters are involved in different channels of the nucleon decay, but not to the $K^+\bar{\nu}$ and $\pi^+\bar{\nu}$ channels. 
With the development of precise experimental measurement in the future, we expect a multi-channel analysis will potentially provide extra information of the flavour space of GUTs that cannot be obtained in the traditional measurements for quark and lepton flavours.

\section*{Acknowledgement}

This work is supported by National Natural Science Foundation of China (NSFC) under Grants Nos. 12205064, 12347103 and Zhejiang Provincial Natural Science Foundation of China under Grant No. LDQ24A050002.


\begin{thebibliography}{99}

%\cite{Georgi:1974sy}
\bibitem{Georgi:1974sy}
H.~Georgi and S.~L.~Glashow,
%``Unity of All Elementary Particle Forces,''
Phys. Rev. Lett. \textbf{32} (1974), 438-441
doi:10.1103/PhysRevLett.32.438
%5755 citations counted in INSPIRE as of 10 Jun 2024

%\cite{Pati:1973rp}
\bibitem{Pati:1973rp}
J.~C.~Pati and A.~Salam,
%``Is Baryon Number Conserved?,''
Phys. Rev. Lett. \textbf{31} (1973), 661-664
doi:10.1103/PhysRevLett.31.661
%905 citations counted in INSPIRE as of 10 Jun 2024

%\cite{Georgi:1974yf}
\bibitem{Georgi:1974yf}
H.~Georgi, H.~R.~Quinn and S.~Weinberg,
%``Hierarchy of Interactions in Unified Gauge Theories,''
Phys. Rev. Lett. \textbf{33} (1974), 451-454
doi:10.1103/PhysRevLett.33.451
%2022 citations counted in INSPIRE as of 04 Jun 2024

%\cite{Ellis:1979fg}
\bibitem{Ellis:1979fg}
J.~R.~Ellis and M.~K.~Gaillard,
%``Fermion Masses and Higgs Representations in SU(5),''
Phys. Lett. B \textbf{88} (1979), 315-319
doi:10.1016/0370-2693(79)90476-3
%270 citations counted in INSPIRE as of 09 Apr 2024

%\cite{Weinberg:1979sa}
\bibitem{Weinberg:1979sa}
S.~Weinberg,
%``Baryon and Lepton Nonconserving Processes,''
Phys. Rev. Lett. \textbf{43} (1979), 1566-1570
doi:10.1103/PhysRevLett.43.1566
%2422 citations counted in INSPIRE as of 10 Jun 2024

%\cite{Shafi:1983gz}
\bibitem{Shafi:1983gz}
Q.~Shafi and C.~Wetterich,
%``Modification of {GUT} Predictions in the Presence of Spontaneous Compactification,''
Phys. Rev. Lett. \textbf{52} (1984), 875
doi:10.1103/PhysRevLett.52.875
%152 citations counted in INSPIRE as of 01 Mar 2024

%\cite{Senjanovic:2024uzn}
\bibitem{Senjanovic:2024uzn}
G.~Senjanovi\'c and M.~Zantedeschi,
%``Minimal SU(5) theory on the edge: The importance of being effective,''
Phys. Rev. D \textbf{109} (2024) no.9, 095009
doi:10.1103/PhysRevD.109.095009
[arXiv:2402.19224 [hep-ph]].
%0 citations counted in INSPIRE as of 03 Jun 2024

%\cite{Glashow:1979nm}
\bibitem{Glashow:1979nm}
S.~L.~Glashow,
%``The Future of Elementary Particle Physics,''
NATO Sci. Ser. B \textbf{61} (1980), 687
doi:10.1007/978-1-4684-7197-7\_15
%756 citations counted in INSPIRE as of 06 Jun 2024

%\cite{Senjanovic:1981ff}
\bibitem{Senjanovic:1981ff}
G.~Senjanovic,
%``NEUTRINO MASS AND UNIFICATION,''
BNL-29092.
%0 citations counted in INSPIRE as of 11 May 2024

%\cite{Magg:1980ut}
\bibitem{Magg:1980ut}
M.~Magg and C.~Wetterich,
%``Neutrino Mass Problem and Gauge Hierarchy,''
Phys. Lett. B \textbf{94} (1980), 61-64
doi:10.1016/0370-2693(80)90825-4
%1245 citations counted in INSPIRE as of 05 Jun 2024

%\cite{Mohapatra:1980yp}
\bibitem{Mohapatra:1980yp}
R.~N.~Mohapatra and G.~Senjanovic,
%``Neutrino Masses and Mixings in Gauge Models with Spontaneous Parity Violation,''
Phys. Rev. D \textbf{23} (1981), 165
doi:10.1103/PhysRevD.23.165
%3060 citations counted in INSPIRE as of 05 Jun 2024

%\cite{Lazarides:1980nt}
\bibitem{Lazarides:1980nt}
G.~Lazarides, Q.~Shafi and C.~Wetterich,
%``Proton Lifetime and Fermion Masses in an SO(10) Model,''
Nucl. Phys. B \textbf{181} (1981), 287-300
doi:10.1016/0550-3213(81)90354-0
%1724 citations counted in INSPIRE as of 05 Jun 2024

%\cite{Bajc:2006ia}
\bibitem{Bajc:2006ia}
B.~Bajc and G.~Senjanovic,
%``Seesaw at LHC,''
JHEP \textbf{08} (2007), 014
doi:10.1088/1126-6708/2007/08/014
[arXiv:hep-ph/0612029 [hep-ph]].
%243 citations counted in INSPIRE as of 22 Apr 2024

%\cite{Bajc:2007zf}
\bibitem{Bajc:2007zf}
B.~Bajc, M.~Nemevsek and G.~Senjanovic,
%``Probing seesaw at LHC,''
Phys. Rev. D \textbf{76} (2007), 055011
doi:10.1103/PhysRevD.76.055011
[arXiv:hep-ph/0703080 [hep-ph]].
%197 citations counted in INSPIRE as of 17 May 2024

%\cite{Dorsner:2005fq}
\bibitem{Dorsner:2005fq}
I.~Dorsner and P.~Fileviez Perez,
%``Unification without supersymmetry: Neutrino mass, proton decay and light leptoquarks,''
Nucl. Phys. B \textbf{723} (2005), 53-76
doi:10.1016/j.nuclphysb.2005.06.016
[arXiv:hep-ph/0504276 [hep-ph]].
%172 citations counted in INSPIRE as of 07 Jun 2024

%\cite{Dorsner:2006hw}
\bibitem{Dorsner:2006hw}
I.~Dorsner, P.~Fileviez Perez and G.~Rodrigo,
%``Fermion masses and the UV cutoff of the minimal realistic SU(5),''
Phys. Rev. D \textbf{75} (2007), 125007
doi:10.1103/PhysRevD.75.125007
[arXiv:hep-ph/0607208 [hep-ph]].
%56 citations counted in INSPIRE as of 14 May 2024

%\cite{Dorsner:2007fy}
\bibitem{Dorsner:2007fy}
I.~Dorsner and I.~Mocioiu,
%``Predictions from type II see-saw mechanism in SU(5),''
Nucl. Phys. B \textbf{796} (2008), 123-136
doi:10.1016/j.nuclphysb.2007.12.004
[arXiv:0708.3332 [hep-ph]].
%59 citations counted in INSPIRE as of 02 May 2024

%\cite{DiLuzio:2013dda}
\bibitem{DiLuzio:2013dda}
L.~Di Luzio and L.~Mihaila,
%``Unification scale vs. electroweak-triplet mass in the $SU(5) + 24_F$ model at three loops,''
Phys. Rev. D \textbf{87} (2013), 115025
doi:10.1103/PhysRevD.87.115025
[arXiv:1305.2850 [hep-ph]].
%17 citations counted in INSPIRE as of 22 Apr 2024

%\cite{Dorsner:2014wva}
\bibitem{Dorsner:2014wva}
I.~Dorsner, S.~Fajfer and I.~Mustac,
%``Light vector-like fermions in a minimal SU(5) setup,''
Phys. Rev. D \textbf{89} (2014) no.11, 115004
doi:10.1103/PhysRevD.89.115004
[arXiv:1401.6870 [hep-ph]].
%32 citations counted in INSPIRE as of 09 Jun 2024

%\cite{FileviezPerez:2016sal}
\bibitem{FileviezPerez:2016sal}
P.~Fileviez Perez and C.~Murgui,
%``Renormalizable SU(5) Unification,''
Phys. Rev. D \textbf{94} (2016) no.7, 075014
doi:10.1103/PhysRevD.94.075014
[arXiv:1604.03377 [hep-ph]].
%42 citations counted in INSPIRE as of 22 Apr 2024

%\cite{Hagedorn:2016dze}
\bibitem{Hagedorn:2016dze}
C.~Hagedorn, T.~Ohlsson, S.~Riad and M.~A.~Schmidt,
%``Unification of Gauge Couplings in Radiative Neutrino Mass Models,''
JHEP \textbf{09} (2016), 111
doi:10.1007/JHEP09(2016)111
[arXiv:1605.03986 [hep-ph]].
%35 citations counted in INSPIRE as of 07 Jun 2024

%\cite{Dorsner:2019vgf}
\bibitem{Dorsner:2019vgf}
I.~Dor\v{s}ner and S.~Saad,
%``Towards Minimal $SU(5)$,''
Phys. Rev. D \textbf{101} (2020) no.1, 015009
doi:10.1103/PhysRevD.101.015009
[arXiv:1910.09008 [hep-ph]].
%13 citations counted in INSPIRE as of 03 Jun 2024

%\cite{Antusch:2021yqe}
\bibitem{Antusch:2021yqe}
S.~Antusch and K.~Hinze,
%``Nucleon decay in a minimal non-SUSY GUT with predicted quark-lepton Yukawa ratios,''
Nucl. Phys. B \textbf{976} (2022), 115719
doi:10.1016/j.nuclphysb.2022.115719
[arXiv:2108.08080 [hep-ph]].
%8 citations counted in INSPIRE as of 17 Jun 2024

%\cite{Antusch:2022afk}
\bibitem{Antusch:2022afk}
S.~Antusch, K.~Hinze and S.~Saad,
%``Viable quark-lepton Yukawa ratios and nucleon decay predictions in SU(5) GUTs with type-II seesaw,''
Nucl. Phys. B \textbf{986} (2023), 116049
doi:10.1016/j.nuclphysb.2022.116049
[arXiv:2205.01120 [hep-ph]].
%10 citations counted in INSPIRE as of 17 Jun 2024

%\cite{Antusch:2023kli}
\bibitem{Antusch:2023kli}
S.~Antusch, K.~Hinze and S.~Saad,
%``Quark-lepton Yukawa ratios and nucleon decay in SU(5) GUTs with type-III seesaw,''
Nucl. Phys. B \textbf{991} (2023), 116195
doi:10.1016/j.nuclphysb.2023.116195
[arXiv:2301.03601 [hep-ph]].
%6 citations counted in INSPIRE as of 17 Jun 2024

%\cite{Antusch:2023mqe}
\bibitem{Antusch:2023mqe}
S.~Antusch, K.~Hinze and S.~Saad,
%``Minimal SU(5) GUTs with vectorlike fermions,''
Phys. Rev. D \textbf{108} (2023) no.9, 095010
doi:10.1103/PhysRevD.108.095010
[arXiv:2308.08585 [hep-ph]].
%5 citations counted in INSPIRE as of 17 Jun 2024

%\cite{Super-Kamiokande:2020wjk}
\bibitem{Super-Kamiokande:2020wjk}
A.~Takenaka \textit{et al.} [Super-Kamiokande],
%``Search for proton decay via $p\to e^+\pi^0$ and $p\to \mu^+\pi^0$ with an enlarged fiducial volume in Super-Kamiokande I-IV,''
Phys. Rev. D \textbf{102} (2020) no.11, 112011
doi:10.1103/PhysRevD.102.112011
[arXiv:2010.16098 [hep-ex]].
%121 citations counted in INSPIRE as of 07 Jun 2024

%\cite{Takhistov:2016eqm}
\bibitem{Takhistov:2016eqm}
V.~Takhistov [Super-Kamiokande],
%``Review of Nucleon Decay Searches at Super-Kamiokande,''
[arXiv:1605.03235 [hep-ex]].
%65 citations counted in INSPIRE as of 08 Jun 2024

%\cite{JUNO:2022qgr}
\bibitem{JUNO:2022qgr}
A.~Abusleme \textit{et al.} [JUNO],
%``JUNO Sensitivity on Proton Decay $p\to \bar\nu K^+$ Searches,''
Chin. Phys. C \textbf{47} (2023) no.11, 113002
doi:10.1088/1674-1137/ace9c6
[arXiv:2212.08502 [hep-ex]].
%20 citations counted in INSPIRE as of 04 Jun 2024

%\cite{DUNE:2020ypp}
\bibitem{DUNE:2020ypp}
B.~Abi \textit{et al.} [DUNE],
%``Deep Underground Neutrino Experiment (DUNE), Far Detector Technical Design Report, Volume II: DUNE Physics,''
[arXiv:2002.03005 [hep-ex]].
%542 citations counted in INSPIRE as of 09 Jun 2024

%\cite{Hyper-Kamiokande:2018ofw}
\bibitem{Hyper-Kamiokande:2018ofw}
K.~Abe \textit{et al.} [Hyper-Kamiokande],
%``Hyper-Kamiokande Design Report,''
[arXiv:1805.04163 [physics.ins-det]].
%885 citations counted in INSPIRE as of 09 Jun 2024

%\cite{Dev:2022jbf}
\bibitem{Dev:2022jbf}
P.~S.~B.~Dev, L.~W.~Koerner, S.~Saad, S.~Antusch, M.~Askins, K.~S.~Babu, J.~L.~Barrow, J.~Chakrabortty, A.~de Gouv\^ea and Z.~Djurcic, \textit{et al.}
%``Searches for baryon number violation in neutrino experiments: a white paper,''
J. Phys. G \textbf{51} (2024) no.3, 033001
doi:10.1088/1361-6471/ad1658
[arXiv:2203.08771 [hep-ex]].
%28 citations counted in INSPIRE as of 08 Jun 2024

%\cite{FileviezPerez:2007bcw}
\bibitem{FileviezPerez:2007bcw}
P.~Fileviez Perez,
%``Renormalizable adjoint SU(5),''
Phys. Lett. B \textbf{654} (2007), 189-193
doi:10.1016/j.physletb.2007.07.075
[arXiv:hep-ph/0702287 [hep-ph]].
%122 citations counted in INSPIRE as of 28 Aug 2024

%\cite{FileviezPerez:2008afb}
\bibitem{FileviezPerez:2008afb}
P.~Fileviez Perez, H.~Iminniyaz and G.~Rodrigo,
%``Proton Stability, Dark Matter and Light Color Octet Scalars in Adjoint SU(5) Unification,''
Phys. Rev. D \textbf{78} (2008), 015013
doi:10.1103/PhysRevD.78.015013
[arXiv:0803.4156 [hep-ph]].
%68 citations counted in INSPIRE as of 28 Aug 2024

%\cite{King:2021gmj}
\bibitem{King:2021gmj}
S.~F.~King, S.~Pascoli, J.~Turner and Y.~L.~Zhou,
%``Confronting SO(10) GUTs with proton decay and gravitational waves,''
JHEP \textbf{10} (2021), 225
doi:10.1007/JHEP10(2021)225
[arXiv:2106.15634 [hep-ph]].
%40 citations counted in INSPIRE as of 04 Jun 2024

%\cite{ATLAS:2022yhd}
\bibitem{ATLAS:2022yhd}
G.~Aad \textit{et al.} [ATLAS],
%``Search for type-III seesaw heavy leptons in leptonic final states in pp collisions at $\sqrt{s} = 13~\text {TeV}$ with the ATLAS detector,''
Eur. Phys. J. C \textbf{82} (2022) no.11, 988
doi:10.1140/epjc/s10052-022-10785-0
[arXiv:2202.02039 [hep-ex]].
%25 citations counted in INSPIRE as of 05 Jun 2024

%\cite{Antusch:2013jca}
\bibitem{Antusch:2013jca}
S.~Antusch and V.~Maurer,
%``Running quark and lepton parameters at various scales,''
JHEP \textbf{11} (2013), 115
doi:10.1007/JHEP11(2013)115
[arXiv:1306.6879 [hep-ph]].
%196 citations counted in INSPIRE as of 05 Jun 2024

%\cite{Babu:2016bmy}
\bibitem{Babu:2016bmy}
K.~S.~Babu, B.~Bajc and S.~Saad,
%``Yukawa Sector of Minimal SO(10) Unification,''
JHEP \textbf{02} (2017), 136
doi:10.1007/JHEP02(2017)136
[arXiv:1612.04329 [hep-ph]].
%74 citations counted in INSPIRE as of 25 Apr 2024

%\cite{Esteban:2020cvm}
\bibitem{Esteban:2020cvm}
I.~Esteban, M.~C.~Gonzalez-Garcia, M.~Maltoni, T.~Schwetz and A.~Zhou,
%``The fate of hints: updated global analysis of three-flavor neutrino oscillations,''
JHEP \textbf{09} (2020), 178
doi:10.1007/JHEP09(2020)178
[arXiv:2007.14792 [hep-ph]].
%1161 citations counted in INSPIRE as of 10 Jun 2024

\bibitem{nufit5.3}
NuFIT 5.3 (2024), http://www.nu-fit.org/?q=node/278 

%\cite{Casas:2001sr}
\bibitem{Casas:2001sr}
J.~A.~Casas and A.~Ibarra,
%``Oscillating neutrinos and $\mu \to e, \gamma$,''
Nucl. Phys. B \textbf{618} (2001), 171-204
doi:10.1016/S0550-3213(01)00475-8
[arXiv:hep-ph/0103065 [hep-ph]].
%1390 citations counted in INSPIRE as of 30 May 2024

%\cite{Ibarra:2003up}
\bibitem{Ibarra:2003up}
A.~Ibarra and G.~G.~Ross,
%``Neutrino phenomenology: The Case of two right-handed neutrinos,''
Phys. Lett. B \textbf{591} (2004), 285-296
doi:10.1016/j.physletb.2004.04.037
[arXiv:hep-ph/0312138 [hep-ph]].
%258 citations counted in INSPIRE as of 04 Jun 2024

%\cite{Arhrib:2009xf}
\bibitem{Arhrib:2009xf}
A.~Arhrib, R.~Benbrik and C.~H.~Chen,
%``Lepton flavor violating tau decays in type-III seesaw mechanism,''
Phys. Rev. D \textbf{81} (2010), 113003
doi:10.1103/PhysRevD.81.113003
[arXiv:0903.1553 [hep-ph]].
%18 citations counted in INSPIRE as of 08 Jun 2024

%\cite{FileviezPerez:2004hn}
\bibitem{FileviezPerez:2004hn}
P.~Fileviez Perez,
%``Fermion mixings versus d = 6 proton decay,''
Phys. Lett. B \textbf{595} (2004), 476-483
doi:10.1016/j.physletb.2004.06.061
[arXiv:hep-ph/0403286 [hep-ph]].
%74 citations counted in INSPIRE as of 20 May 2024

%\cite{FileviezPerez:2018dyf}
\bibitem{FileviezPerez:2018dyf}
P.~Fileviez P\'erez, A.~Gross and C.~Murgui,
%``Seesaw scale, unification, and proton decay,''
Phys. Rev. D \textbf{98} (2018) no.3, 035032
doi:10.1103/PhysRevD.98.035032
[arXiv:1804.07831 [hep-ph]].
%14 citations counted in INSPIRE as of 22 Apr 2024

%\cite{Nath:2006ut}
\bibitem{Nath:2006ut}
P.~Nath and P.~Fileviez Perez,
%``Proton stability in grand unified theories, in strings and in branes,''
Phys. Rept. \textbf{441} (2007), 191-317
doi:10.1016/j.physrep.2007.02.010
[arXiv:hep-ph/0601023 [hep-ph]].
%498 citations counted in INSPIRE as of 30 May 2024

%\cite{Aoki:2017puj}
\bibitem{Aoki:2017puj}
Y.~Aoki, T.~Izubuchi, E.~Shintani and A.~Soni,
%``Improved lattice computation of proton decay matrix elements,''
Phys. Rev. D \textbf{96} (2017) no.1, 014506
doi:10.1103/PhysRevD.96.014506
[arXiv:1705.01338 [hep-lat]].
%136 citations counted in INSPIRE as of 04 Jun 2024

%\cite{Workman:2022ynf}
\bibitem{Workman:2022ynf}
R.~L.~Workman \textit{et al.} [Particle Data Group],
%``Review of Particle Physics,''
PTEP \textbf{2022} (2022), 083C01
doi:10.1093/ptep/ptac097
%3260 citations counted in INSPIRE as of 10 Jun 2024

%\cite{Super-Kamiokande:2022egr}
\bibitem{Super-Kamiokande:2022egr}
R.~Matsumoto \textit{et al.} [Super-Kamiokande],
%``Search for proton decay via $p\rightarrow \mu^+K^0$ in 0.37 megaton-years exposure of Super-Kamiokande,''
Phys. Rev. D \textbf{106} (2022) no.7, 072003
doi:10.1103/PhysRevD.106.072003
[arXiv:2208.13188 [hep-ex]].
%22 citations counted in INSPIRE as of 07 Jun 2024

%\cite{Super-Kamiokande:2013rwg}
\bibitem{Super-Kamiokande:2013rwg}
K.~Abe \textit{et al.} [Super-Kamiokande],
%``Search for Nucleon Decay via $n \to \bar{\nu} \pi^{0}$ and $p \to \bar{\nu} \pi^{+}$ in Super-Kamiokande,''
Phys. Rev. Lett. \textbf{113} (2014) no.12, 121802
doi:10.1103/PhysRevLett.113.121802
[arXiv:1305.4391 [hep-ex]].
%95 citations counted in INSPIRE as of 08 Jun 2024

%\cite{Yokoyama:2017mnt}
\bibitem{Yokoyama:2017mnt}
M.~Yokoyama [Hyper-Kamiokande Proto],
%``The Hyper-Kamiokande Experiment,''
[arXiv:1705.00306 [hep-ex]].
%34 citations counted in INSPIRE as of 23 May 2024

%\cite{Super-Kamiokande:2005lev}
\bibitem{Super-Kamiokande:2005lev}
K.~Kobayashi \textit{et al.} [Super-Kamiokande],
%``Search for nucleon decay via modes favored by supersymmetric grand unification models in Super-Kamiokande-I,''
Phys. Rev. D \textbf{72} (2005), 052007
doi:10.1103/PhysRevD.72.052007
[arXiv:hep-ex/0502026 [hep-ex]].
%178 citations counted in INSPIRE as of 08 Jun 2024
\end{thebibliography}
\end{document}